\documentclass[twocolumn]{aastex63}

\usepackage{graphicx}
\usepackage{natbib}
\usepackage{placeins}
\usepackage{amsmath}
\usepackage{tgcursor}
\usepackage{booktabs}
\usepackage{enumitem}
\setcitestyle{notesep={ }}
\usepackage{booktabs}
\usepackage{multirow} 
\usepackage{rotating}

\newcommand{\hh}{H$_2$}
\newcommand{\cch}{C$_2$H}
\newcommand{\hhco}{H$_2$CO}
\newcommand{\twco}{$^{12}$CO}
\newcommand{\thco}{$^{13}$CO}

\newcommand{\Tex}{\rm{T}_{\rm{ex}}}
\newcommand{\TexO}{\rm{T}_{\rm{ex},0}}
\newcommand{\Tbg}{\rm{T}_{\rm{bg}}}

\newcommand{\Tgas}{\rm{T}_{\rm{gas}}}

\newcommand{\pccm}{~\rm{cm}^{-3}}
\newcommand{\pscm}{~\rm{cm}^{-2}}
\newcommand{\kms}{~\rm{km~s}^{-1}}

\newcommand{\TabDiskSample}{%
    \begin{table}[t!]
    \begin{center}    
     \caption{Source characterization of the MAPS sample.}
        \label{tab:disk_sample}   
        \begin{tabular}{lccccc}\toprule
        Source & Spectral & Distance & Incl & PA & $M_{\star}$\\   
        & Type & (pc) & ($^{\circ}$) & ($^{\circ}$) & ($M_{\odot}$)\\
        \midrule
        IM Lup    & K5 & 158   & 47.5  & 144.5 & 1.1\\
        GM Aur    & K6 & 159   & 53.2  & 57.2  & 1.1\\
        AS 209    & K5 & 121   & 35.0  & 85.8  & 1.2\\
        HD 162396 & A1 & 101   & 46.7  & 133.3 & 2.0\\
        MWC 480   & A5 & 162 & 37.0    & 148.0   & 2.1\\
        \bottomrule
        \end{tabular}
    \end{center}
    Reproduced from \cite{oberg21}, where additional source characteristics and a list of references are provided.
    \end{table}
}

\newcommand{\TabSpecParams}{%
  \begin{table*}
    \begin{center}
    \caption{Spectroscopic constants of the observed lines.}
    \label{tab:spec_params}
        \begin{tabular}{ccccccccc}\toprule
            Molecule & Transition &  $\nu$ & $S_{ij} \mu^2$  & $\log_{10}(A_{ij})$  & E$_{\rm{u}}$ & $g_u$  & $v_i\tablenotemark{a}$ & $r_i\tablenotemark{b}$ \\
            & & (GHz) & ($D^2$) & ($s^{-1}$) & (K) & & ($\kms$) &\\
            \midrule
            \multirow{9}{*}{HCN}  
            & $J=1-0$, $F=1-1$ & 88.630415 	& 8.91247  & $-4.61844$ & 	4.2  & 3 & 4.8 & 0.600 \\	
            & $J=1-0$, $F=2-1$ & 88.631847 	& 14.85197 & $-4.61848$ & 	4.2  & 5 & 0.0 & 1.000	\\
            & $J=1-0$, $F=0-1$ & 88.633935 	& 2.97073  & $-4.61840$ & 	4.2  & 	1 &	$-7.1$ & 0.200\\ [0.6ex]
            & $J=3-2$, $F=3-3$ & 265.884891 	& 2.97015  & $-4.03231$ & 	25.5 & 	7 &	$1.8$ & 0.086 \\
            & $J=3-2$, $F=2-1$ & 265.886188 	& 16.03887 & $-3.15378$ & 	25.5 & 	5 &	$0.4$ & 0.467\\
            & $J=3-2$, $F=3-2$ & 265.886433 	& 23.76144 & $-3.12920$ & 	25.5 & 	7 &	$0.1$ & 0.691 \\
            & $J=3-2$, $F=4-3$ & 265.886499 	& 34.36945 & $-3.07805$ & 	25.5 & 	9 &	$0.0$ & 1.000\\
            & $J=3-2$, $F=2-3$ & 265.886979 	& 0.08487  & $-5.43017$ & 	25.5 & 	5 &	$-0.5$ & 0.002\\
            & $J=3-2$, $F=2-2$ & 265.888522 	& 2.97075  & $-3.88607$ & 	25.5 & 	5 &	$-2.3$ & 0.086\\
            \midrule
            \multirow{6}{*}{\cch{}} 
            & $N=1-0$, $J=3/2-1/2$, $F=2-1$ & 87.316925   & 1.42458 & $-5.65605$ & 4.2 & 5 & 0.0 & 1.000	\\	
            & $N=1-0$, $J=3/2-1/2$, $F=1-0$ & 87.328624  & 0.70952 & $-5.73675$ & 4.2 & 	3 & $-40.2$ & 0.498\\	[0.6ex]
            & $N=3-2$, $J=7/2-5/2$, $F=4-3$ & 262.004226 & 3.29692 & $-4.11528$ & 25.1 & 	9 & 0.0 & 1.000\\	
            & $N=3-2$, $J=7/2-5/2$, $F=3-2$ & 262.006403 & 2.46660 & $-4.13213$ & 25.1 & 	7 & $-2.5$ & 0.748\\	
            & $N=3-2$, $J=5/2-3/2$, $F=3-2$ & 262.064843 & 2.35407 & $-4.15212$ & 25.2 & 	7 & $-69.4$ & 0.714\\	
            & $N=3-2$, $J=5/2-3/2$, $F=2-1$ & 262.067331 & 1.53855 & $-4.19069$ & 25.2 & 	5 & $-72.2$ & 0.467\\
            & $N=3-2$, $J=5/2-3/2$, $F=2-2$ & 262.078935 & 0.20691 & $-5.06197$ & 25.2 &	5 &	$-85.5$ & 0.063\\
            \midrule
            \hhco{} & $3_{03}-2_{02}$ & 218.222192 &	16.29674	& $-3.55037$	& 20.95640	& 7 & 0 & 1.000\\	 
          \bottomrule
        \end{tabular}
    \end{center}
    The spectroscopic constants are taken from the CDMS database \citep{cdmsA,cdmsB} for HCN, and from the JPL database \citep{jpl} for \cch{} and \hhco{}. Measurements are provided by \citet{Ahrens2002} for HCN, by \citet{Sastry1981} and \citet{Padovani2009} for \cch{}, and by \citet{Bocquet1996} for \hhco{}.
    \tablenotetext{a}{Velocity shift from the brightest hyperfine component in the group.} \tablenotetext{b}{Relative intensity with respect to the brightest hyperfine component in the group.
    }
\end{table*}
}

\newcommand{\TabFluxes}{%
\begin{sidewaystable}
    \begin{center}
    \caption{HCN, \cch{}, and \hhco{} Disk-integrated lines fluxes.}
    \label{tab:fluxes}
        \begin{tabular}{clccccc}\toprule
            Molecule & Line & IM~Lup & GM~Aur & AS~209 & HD~163296 & MWC~480 \\
            & & (mJy km s$^{-1}$) & (mJy km s$^{-1}$) & (mJy km s$^{-1}$) & (mJy km s$^{-1}$) & (mJy km s$^{-1}$) \\
            \midrule
            \multirow{2}{*}{HCN}  
            & $J=1-0$, $F=1-1; F=2-1; F=0-1$ & 88.3$\pm$  14.4 &	  113.4$\pm$  14.8 &	  226.8$\pm$  15.9 &	  736.3$\pm$  58.3 &	  138.6$\pm$  19.6\\
            & $J=3-2$, $F=3-2; F=3-3; F=2-2$ & 2343.7$\pm$  86.3 & 	 1814.6$\pm$ 130.5 &	 2940.8$\pm$  71.7 &	 7043.3$\pm$ 459.3 &	 2494.4$\pm$ 122.8\\
            \midrule
            \multirow{5}{*}{\cch{}} 
            & $N=1-0$, $J=3/2-1/2$, $F=2-1$ & $<$30           &	   $<$16           &	   60.7$\pm$   9.7 &	  130.1$\pm$  34.2 &	   44.3$\pm$  17.2\\
            & $N=1-0$, $J=3/2-1/2$, $F=1-0$ & $<$8            &	   $<$13.9         &	   40.1$\pm$  11.9 &	   56.1$\pm$  14.1 &	   $<$15.1\\
            & $N=3-2$, $J=7/2-5/2$, $F=4-3; F=3-2$ & 430.8$\pm$  59.3 & 	  294.2$\pm$  97.9 &	 1716.6$\pm$  56.7 &	 2527.5$\pm$ 223.1 &	 1076.4$\pm$  66.6\\
            & $N=3-2$, $J=5/2-3/2$, $F=3-2; F=2-1$ & 620.2$\pm$  88.3 &	  506.7$\pm$ 102.0 &	 2119.7$\pm$  73.9 &	 3230.6$\pm$ 400.4 &	 1418.0$\pm$  72.7\\
            & $N=3-2$, $J=5/2-3/2$, $F=2-2$ & $<$37           &	   $<$77           &	  142.4$\pm$  33.8 &	  188.4$\pm$  69.4 & 36.3$\pm$  18.2\\
            \midrule
            \hhco{} & $3_{03}-2_{02}$ & 752.2$\pm$  72.1 &	  900.0$\pm$  56.7 &	  292.5$\pm$  27.9 &	  813.4$\pm$  78.1 &	  144.5$\pm$  38.9\\
          \bottomrule
        \end{tabular}
    \end{center}
\end{sidewaystable}
}

\newcommand{\TabOrganicreservoir}{%
    \begin{table*}[t!]
        \begin{center}
       \caption{Small organic reservoir in the inner $50$~au.} 
        \label{tab:organic_reservoir}   
        \begin{tabular}{lccc|ccc}\toprule
        & \multicolumn{3}{c|}{Gas Mass} & \multicolumn{3}{c}{Gas$+$ice mass w.r.t. H$_2$O ice\tablenotemark{$^\dagger$}} \\  
        Source & HCN & \cch{}  & \hhco{} & HCN & \cch{}  & \hhco{}  \\
            & ($10^{22}$~g) & ($10^{22}$~g) & ($10^{22}$~g) & (\% H$_2$O) & (\% H$_2$O) & (\% H$_2$O) \\    
        \midrule
        IM~Lup 	    &  0.03 &  0.05 &  0.20 	 &  $<0.001$ & $<0.001$ & 0.001\\
        GM~Aur 	    & 19.63 &  2.04 &  0.79 	 &  0.56 &  0.06 &  0.02\\
        AS~209 	    &  1.99 &  3.27 &  0.15 	 &  1.09 &  1.80 &  0.08\\
        HD~163296 	& 46.36 &  6.03 &  0.24 	 &  0.68 &  0.09 &  0.004\\
        MWC~480 	& 59.92 &  2.17 &  1.21 	 &  0.45 &  0.02 &  0.01\\
        \bottomrule
        \end{tabular}\\
    $^\dagger$ Assuming an ice-to-gas ratio of 1000
    \end{center}
    \end{table*}
}

\newcommand{\FigMaps}{%
\begin{figure*}[t!]
    \includegraphics[width=\linewidth]{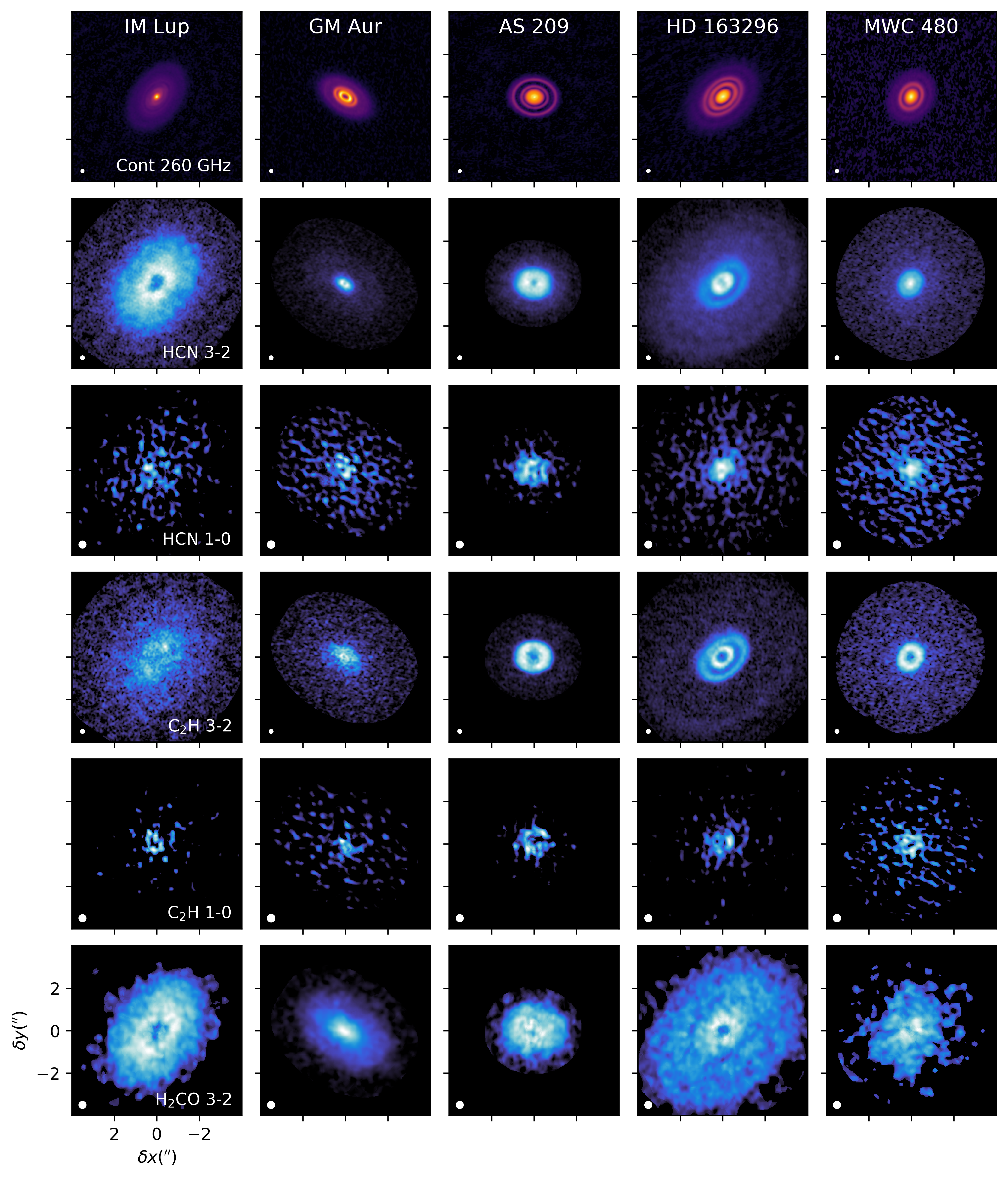}
    \caption{260~GHz dust continuum emission and zeroth-moment maps for the HCN $J=3-2$ and $J=1-0$, \cch{} $N=3-2$, $N=1-0$ and \hhco{} $J=3-2$ lines. All the hyperfine components were included to generate the HCN zeroth-moment map, while only the brightest hyperfine component is included in the \cch{} zeroth-moment map. The color is shown with power-law stretch to enhance the faint emission in the outer disk. The beam is shown in the bottom left corner of each panel.}
    \label{fig:maps}
\end{figure*}
}

\newcommand{\FigProfs}{%
\begin{figure*}[t!]
    \includegraphics[width=\linewidth]{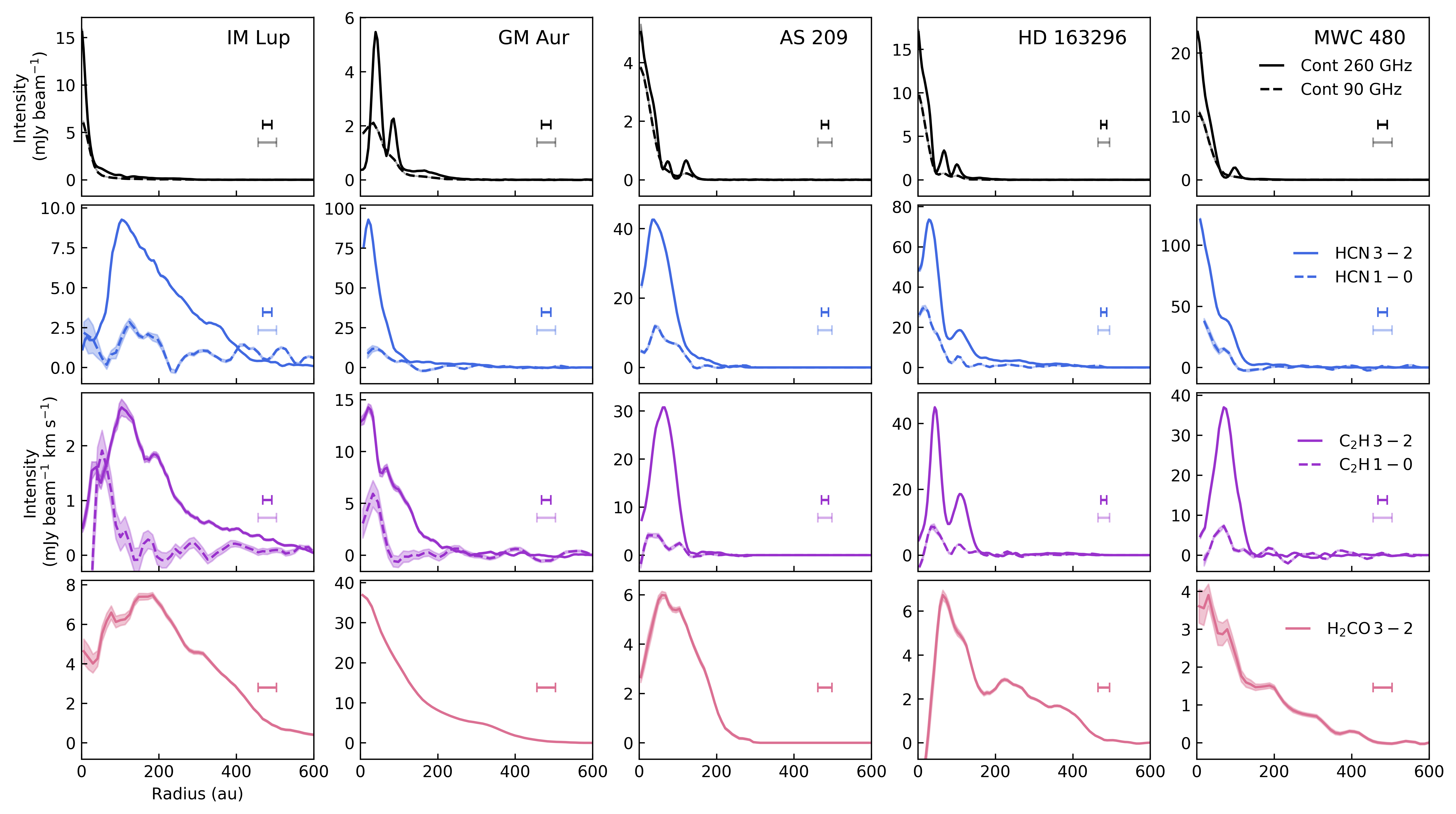}
    \caption{Deprojected radial intensity profiles of the continuum (first row), HCN (second row), \cch{} (third row), and \hhco{} (bottom row) emission lines. Solid and dashed lines correspond to the Band 6 and Band 3 observations, respectively. The horizontal bars mark the beam size of the observations, where light colors correspond to the Band 3 data. Color-shaded regions show the $1\sigma$ scatter at each radial bin divided by the square root of the ratio of bin circumference and FWHM of the synthesized beam \citep[see][]{law21_rad}.}
    \label{fig:profs}
\end{figure*}
}

\newcommand{\FigNprofs}{%
\begin{figure*}
    \includegraphics[width=\linewidth]{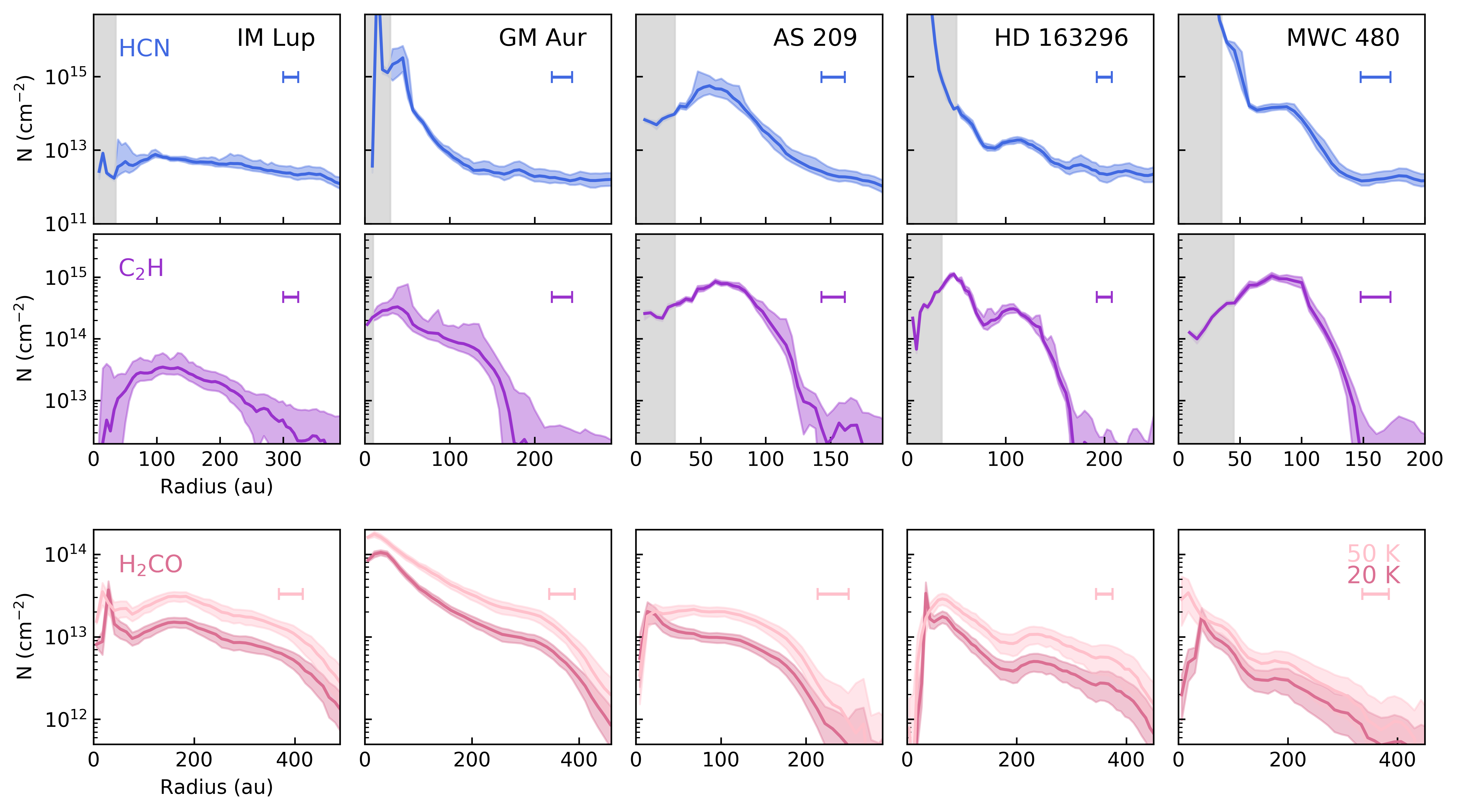}
    \caption{Column density profiles for HCN (top), \cch{} (middle), and \hhco{} (bottom). The horizontal bar at the upper right marks the spatial resolution of the profiles. Gray shaded areas mark the regions where the excitation temperature was assumed to be equal to the \twco{} or \thco{} temperature. Two column density profiles are shown for \hhco{}, assuming a fixed excitation temperature of 20 and 50 K. }
    \label{fig:Nprofs}
\end{figure*}
}

\newcommand{\FigTexprofs}{%
\begin{figure*}
    \includegraphics[width=\linewidth]{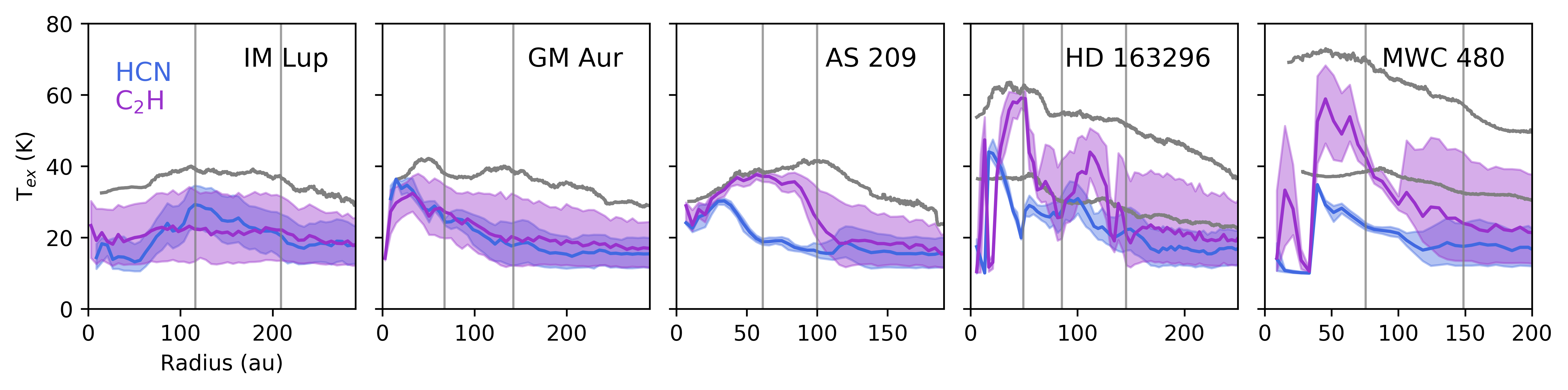}
    \caption{Excitation temperatures for HCN and \cch{} derived from the fit to the $3-2$ lines only. The $\Tgas$(\twco) temperature is shown in gray for all sources. In addition, the $\Tgas$(\thco) temperature is shown for HD~163206 and MWC~480. The horizontal bar at the upper right marks the resolution of the profiles. The gray vertical lines show the position of the gaps seen in the dust millimeter continuum as measured by \citet{law21_rad}.}
    \label{fig:Texprofs}
\end{figure*}
}

\newcommand{\FigNprofsCombined}{%
\begin{figure*}
    \includegraphics[width=\linewidth]{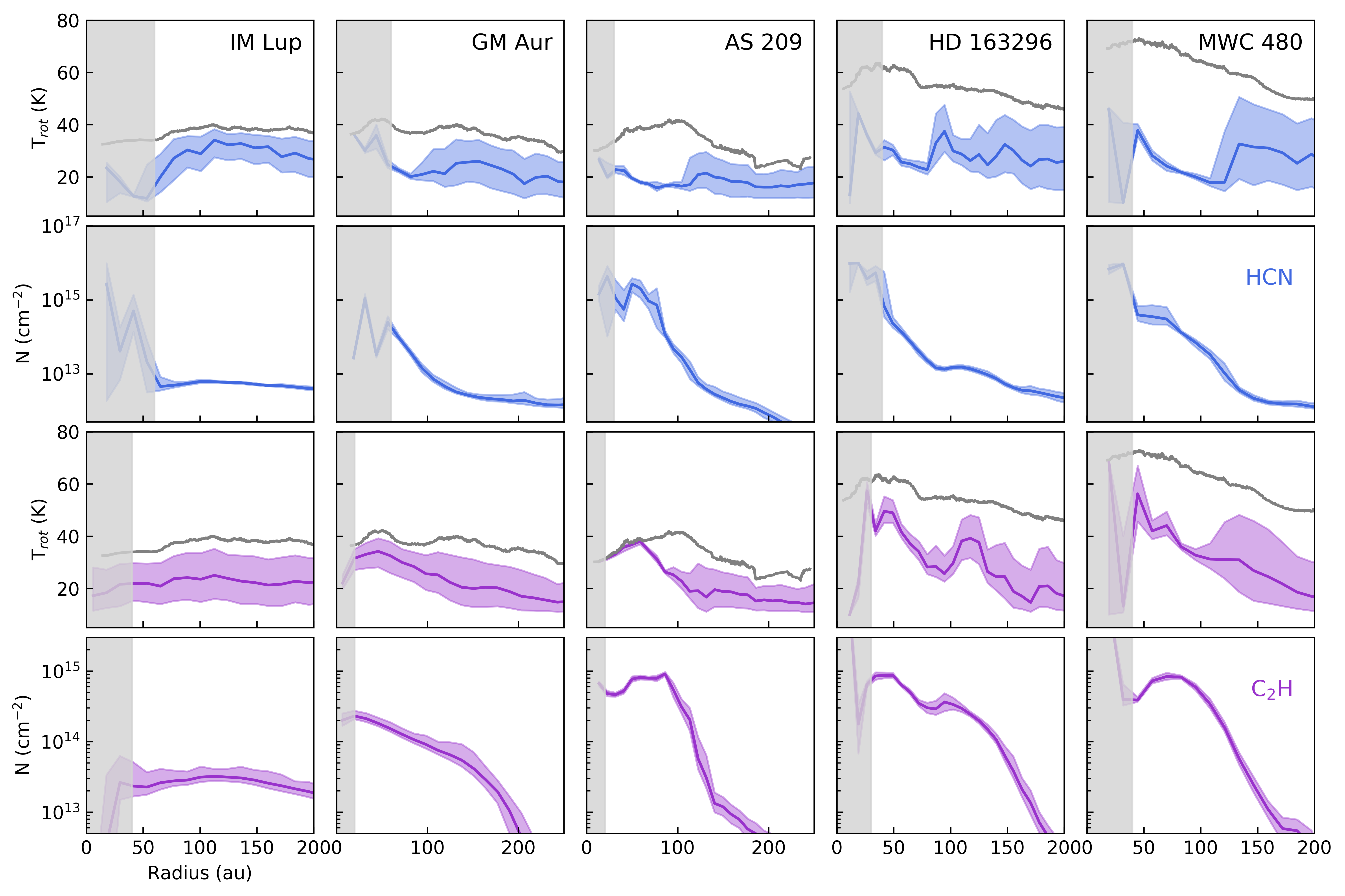}
    \caption{Best-fit rotational temperature and column density profiles to the $1-0$ and $3-2$ lines. The two upper and two bottom rows correspond to HCN and \cch{}, respectively. The gray curves correspond to the gas temperatures derived from the optically thick \twco{} 2--1 line \citep{law21_surf} and was used as the maximum rotational temperature allowed in the fit. Gray shaded areas mark the regions where the lines are poorly fit. The resolution of the profiles is 0\farcs3.}
    \label{fig:NprofsCombined}
\end{figure*}
}

\newcommand{\FigNprofsComparisonCO}{%
\begin{figure*}
    \includegraphics[width=\linewidth]{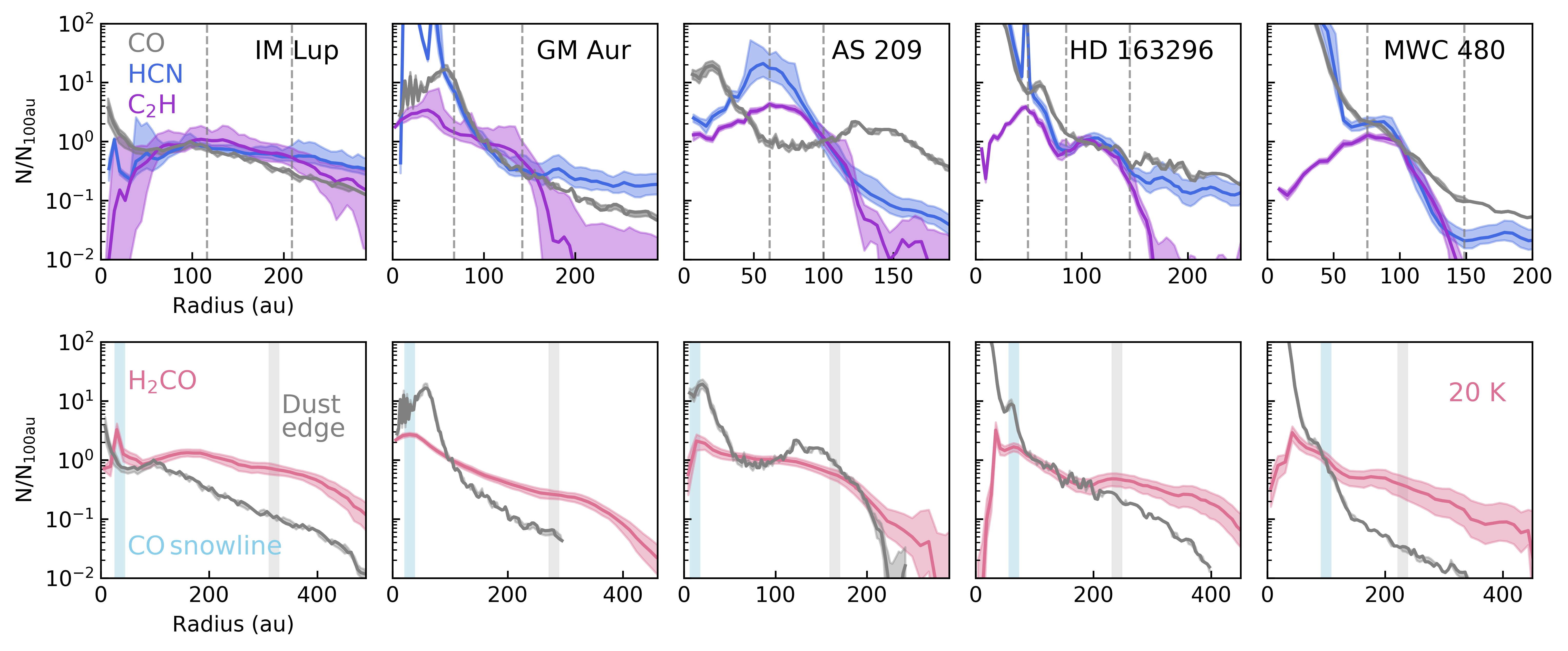}
    \caption{Normalized column density profiles for HCN, \cch{}, (top) and \hhco{} (bottom). For comparison, the CO column density profiles are shown in gray \citep{zhang21}. The vertical dashed lines show the position of the gaps seen in the dust continuum as measured by \citet{law21_rad}. The light blue vertical lines mark the estimated location of the CO snowline \citep{zhang21}. The light gray vertical lines mark the edge of the millimeter dust disk.}
    \label{fig:Nprofs-comp}
\end{figure*}
}

\newcommand{\FigNprofsComparisonHCN}{%
\begin{figure*}
    \includegraphics[width=\linewidth]{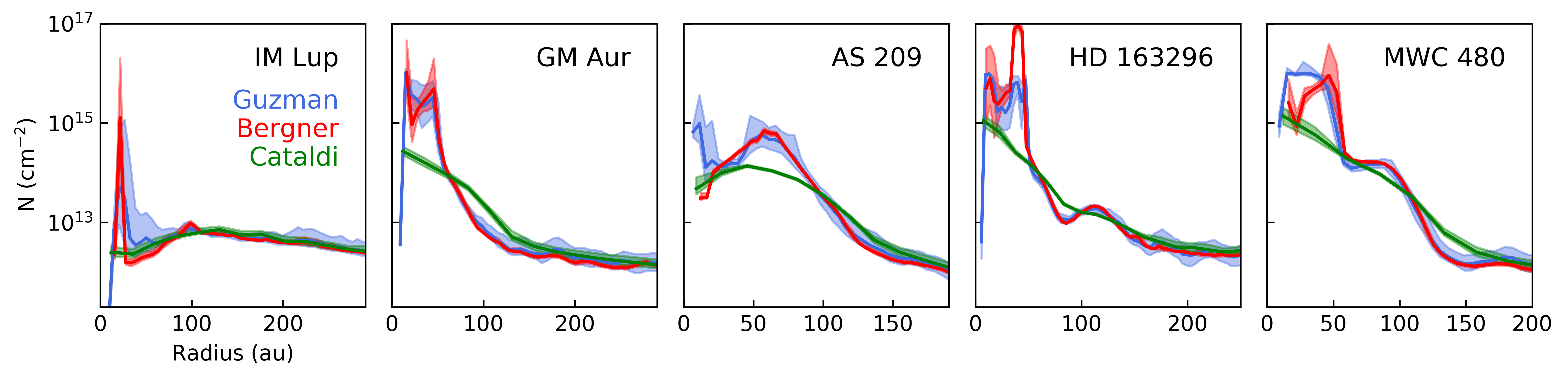}
    \caption{Column density profiles for HCN derived from this work (blue), \citet{bergner21} (red), and \citet{Cataldi21} (green).}
    \label{fig:Nprofs-compHCN}
\end{figure*}
}

\newcommand{\FigSpecFitHighRes}{%
\begin{figure*}[t!]
    \includegraphics[width=\linewidth]{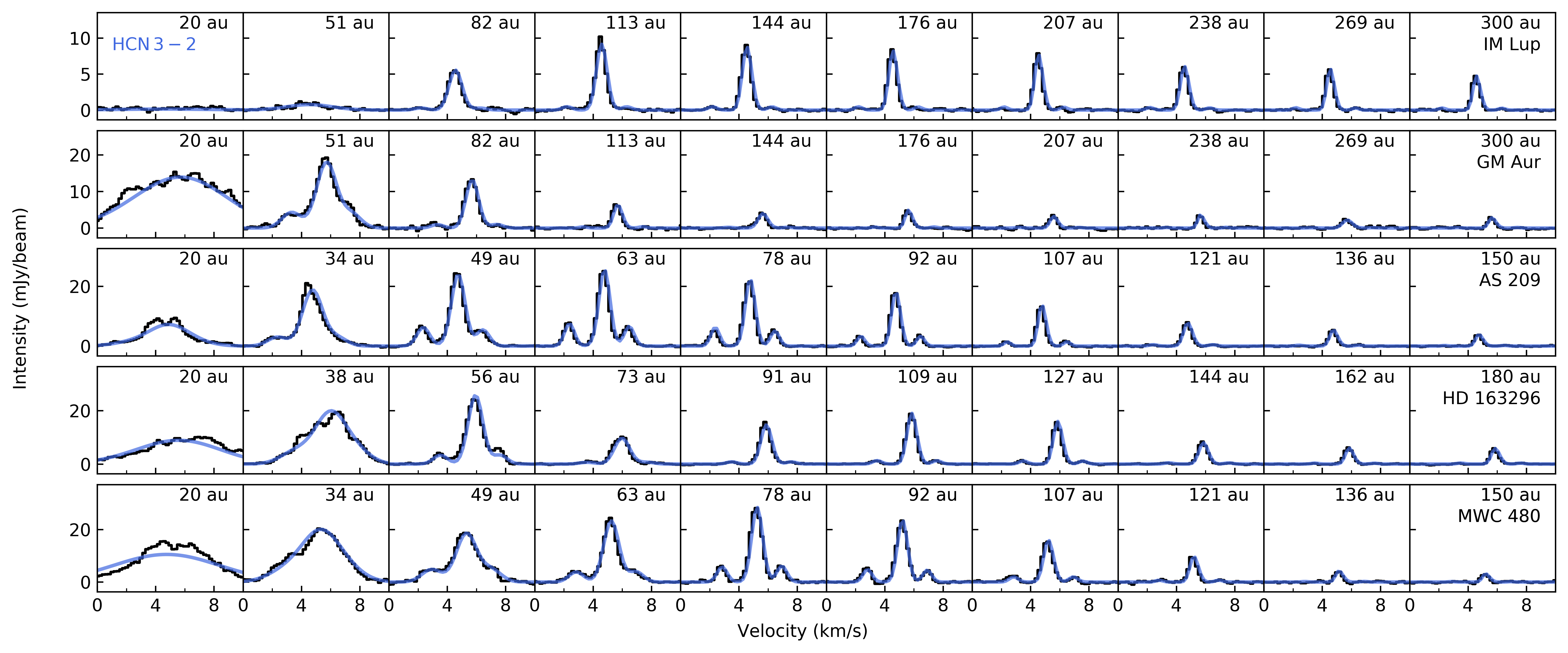}\\
    \includegraphics[width=\linewidth]{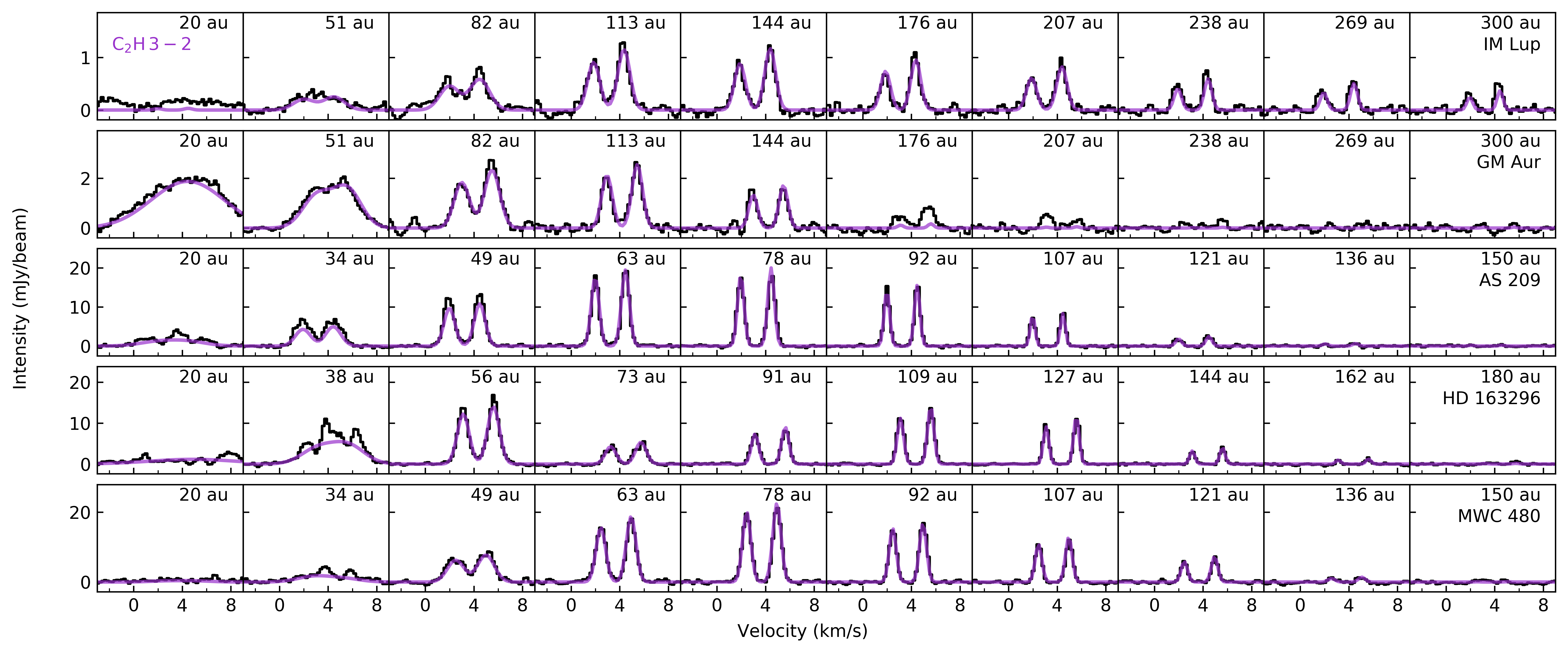}\\
    \includegraphics[width=\linewidth]{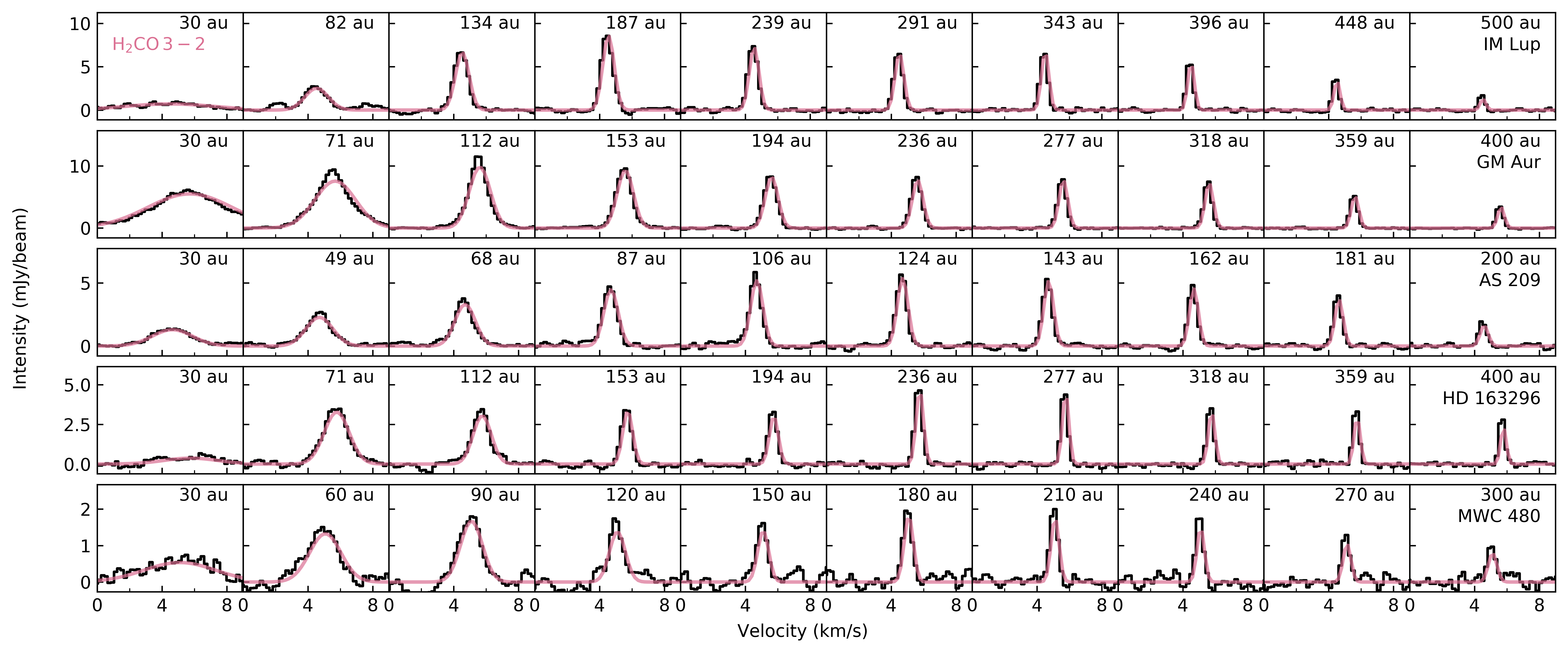}
    \caption{Deprojected spectra averaged over radial annuli for HCN (top), \cch{} (middle), and \hhco{} (bottom) lines. The observed spectra are shown in black, and the best-fit are shown in color. For \hhco{}, the best-fit model for a temperature of 20~K is shown. The radius of the center of each radial bin is shown at the top of each panel. }
    \label{fig:SpecFitB6}
\end{figure*}
}

\newcommand{\FigProfsHD}{%
\begin{figure}[b!]
    \includegraphics[width=\linewidth]{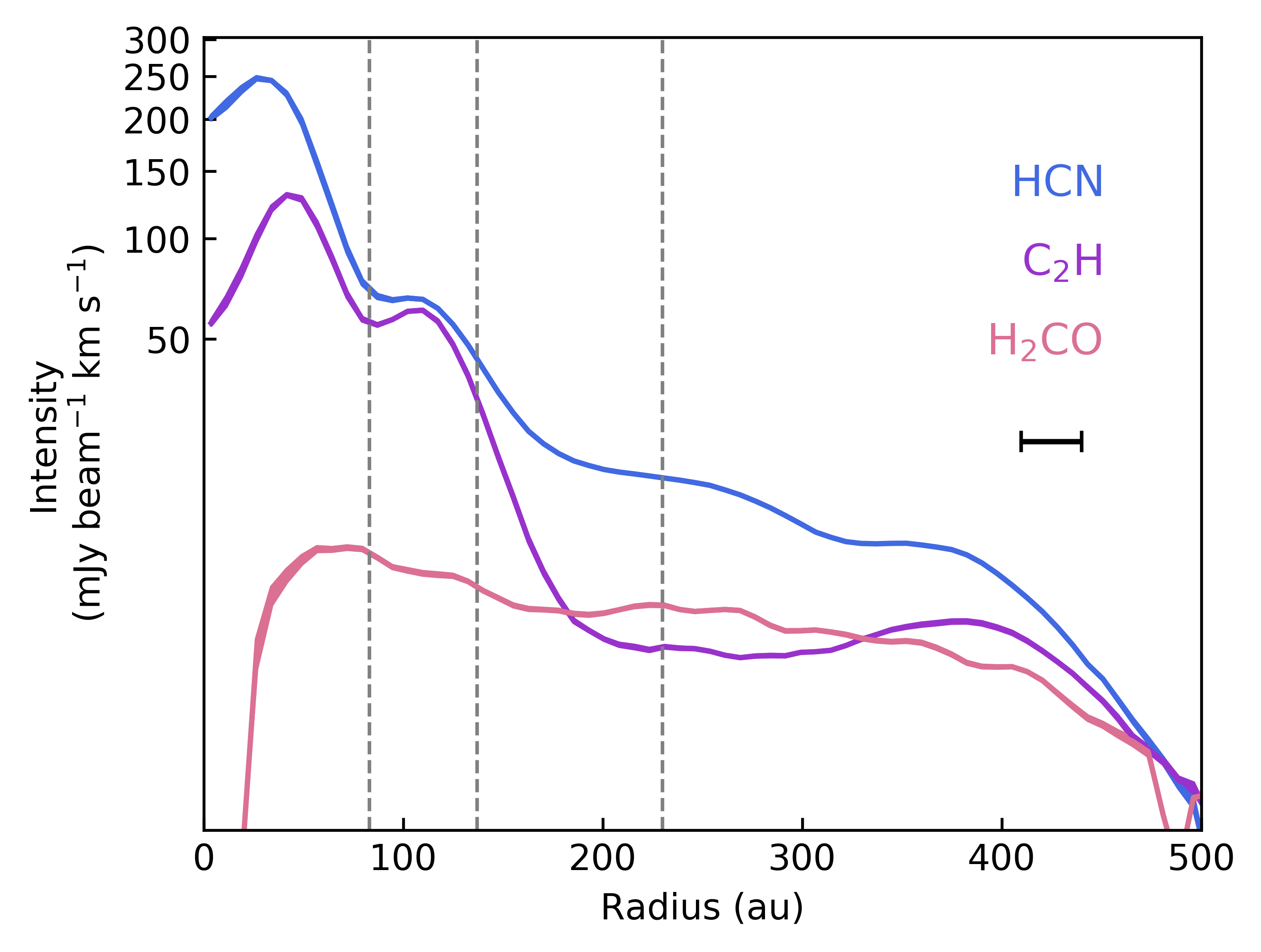}
    \caption{Deprojected radial profiles of the HCN, \cch{}, and \hhco{} $3-2$ lines toward HD~163296. The profiles are computed from the tapered $0\farcs3$ images. The dashed vertical lines mark the position of the putative planets at 83, 137, and 230~au \citep{teague2018b,pinte2018}.}
    \label{fig:ProfHD}
\end{figure}
}

\newcommand{\FigSpecLowResAS}{%
\begin{figure*}
    \includegraphics[width=\linewidth]{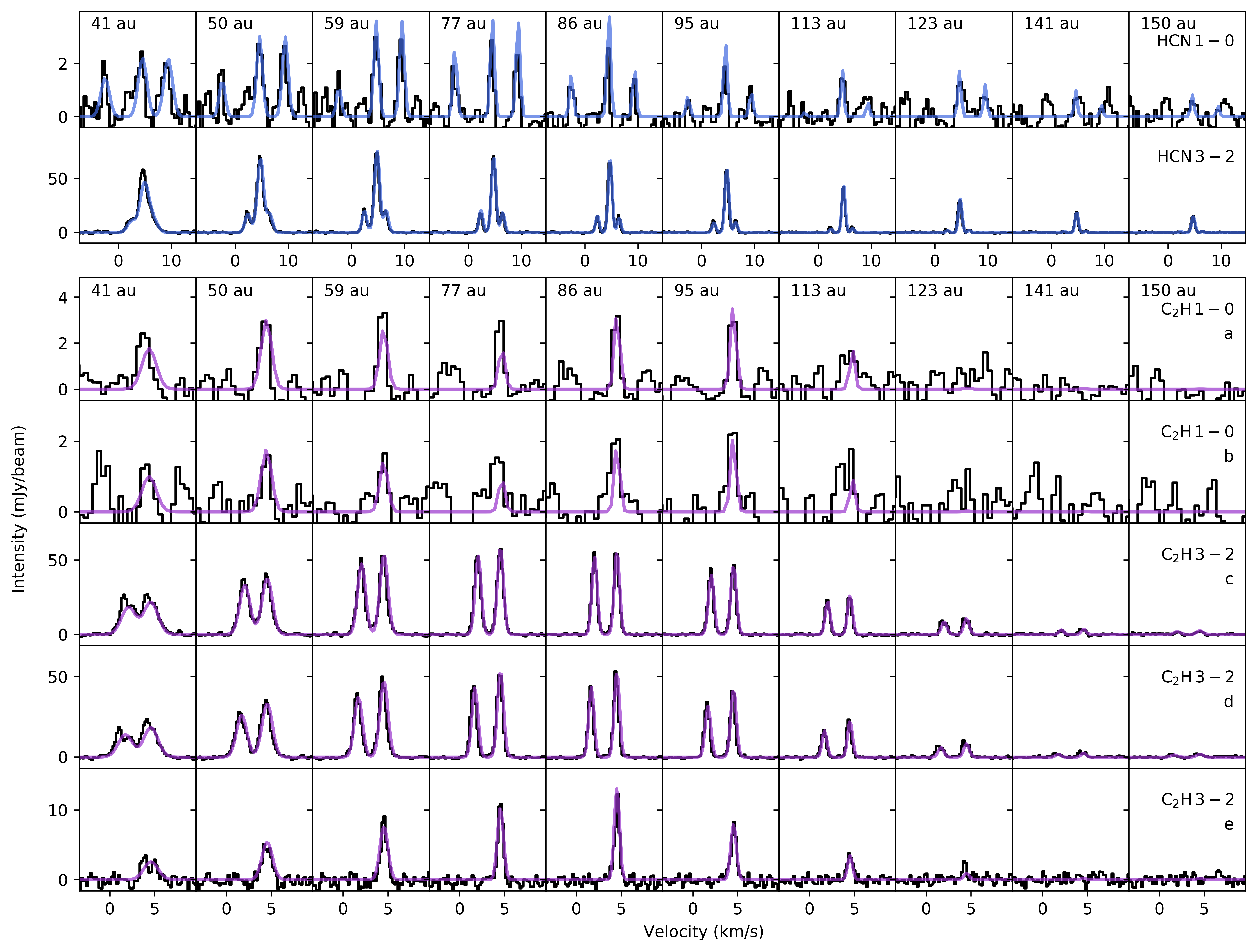}
    \caption{Deprojected spectra averaged over radial annuli in AS~209. HCN and \cch{} lines are shown in the top and bottom panels, respectively. The $3-2$ lines correspond to the tapered images with 0\farcs3 resolution. The observed spectra are shown in black, and the best-fit are shown in color. Several hyperfine lines are shown for \cch{}: (a) and (b) correspond to the $F=2-1$ and $F=1-0$ components of the $1-0$ line; (c) corresponds to the $F=4-3$ and $F=3-2$ components associated with the $J=7/2-5/2$ group of the $3-2$ line; (d) corresponds to the $F=3-2$ and $F=2-1$ components associated with the $J=5/2-3/2$ group of the $3-2$ line; and (e) corresponds to the $F=2-2$ component associated with the $J=5/2-3/2$ group of the $3-2$ line.}
    \label{fig:spec-as209}
\end{figure*}
}

\accepted{July 14, 2021}
\submitjournal{ApJ}

\shorttitle{MAPS VI}
\shortauthors{Guzm\'an et al.}
\graphicspath{{./}{figures/}}

\begin{document}

\title{Molecules with ALMA at Planet-forming Scales (MAPS) VI: \\Distribution of the small organics HCN, C$_2$H, and H$_2$CO}

\correspondingauthor{Viviana V. Guzm\'{a}n}
\email{vguzman@astro.puc.cl}

\author[0000-0003-4784-3040]{Viviana V. Guzm\'{a}n}
\affiliation{Instituto de Astrof\'isica, Pontificia Universidad Cat\'olica de Chile, Av. Vicu\~na Mackenna 4860, 7820436 Macul, Santiago, Chile}

\author[0000-0002-8716-0482]{Jennifer B. Bergner} 
\altaffiliation{NASA Hubble Fellowship Program Sagan Fellow}
\affiliation{University of Chicago Department of the Geophysical Sciences, Chicago, IL 60637, USA}

\author[0000-0003-1413-1776]{Charles J. Law}
\affiliation{Center for Astrophysics \textbar\, Harvard \& Smithsonian, 60 Garden St., Cambridge, MA 02138, USA}

\author[0000-0001-8798-1347]{Karin I. \"Oberg} 
\affiliation{Center for Astrophysics \textbar\, Harvard \& Smithsonian, 60 Garden St., Cambridge, MA 02138, USA}

\author[0000-0001-6078-786X]{Catherine Walsh}
\affiliation{School of Physics and Astronomy, University of Leeds, Leeds, LS2 9JT, UK}

\author[0000-0002-2700-9676]{Gianni Cataldi}
\affiliation{National Astronomical Observatory of Japan, 2-21-1 Osawa, Mitaka, Tokyo 181-8588, Japan}
\affiliation{Department of Astronomy, Graduate School of Science, The University of Tokyo, Tokyo 113-0033, Japan}

\author[0000-0003-3283-6884]{Yuri Aikawa}
\affiliation{Department of Astronomy, Graduate School of Science, The University of Tokyo, Tokyo 113-0033, Japan}

\author[0000-0003-4179-6394]{Edwin A. Bergin}
\affiliation{Department of Astronomy, University of Michigan, 323 West Hall, 1085 S. University Avenue, Ann Arbor, MI 48109, USA}

\author[0000-0002-1483-8811]{Ian Czekala}
\affiliation{Department of Astronomy and Astrophysics, 525 Davey Laboratory, The Pennsylvania State University, University Park, PA 16802, USA}
\affiliation{Center for Exoplanets and Habitable Worlds, 525 Davey Laboratory, The Pennsylvania State University, University Park, PA 16802, USA}
\affiliation{Center for Astrostatistics, 525 Davey Laboratory, The Pennsylvania State University, University Park, PA 16802, USA}
\affiliation{Institute for Computational \& Data Sciences, The Pennsylvania State University, University Park, PA 16802, USA}
\altaffiliation{NASA Hubble Fellowship Program Sagan Fellow}
\affiliation{Department of Astronomy, 501 Campbell Hall, University of California, Berkeley, CA 94720-3411, USA}


\author[0000-0001-6947-6072]{Jane Huang}
\altaffiliation{NASA Hubble Fellowship Program Sagan Fellow}
\affiliation{Department of Astronomy, University of Michigan, 323 West Hall, 1085 S. University Avenue, Ann Arbor, MI 48109, USA}
\affiliation{Center for Astrophysics \textbar\, Harvard \& Smithsonian, 60 Garden St., Cambridge, MA 02138, USA}

\author[0000-0003-2253-2270]{Sean M. Andrews} \affiliation{Center for Astrophysics \textbar\, Harvard \& Smithsonian, 60 Garden St., Cambridge, MA 02138, USA}

\author[0000-0002-8932-1219]{Ryan A. Loomis}
\affiliation{National Radio Astronomy Observatory, 520 Edgemont Rd., Charlottesville, VA 22903, USA}

\author[0000-0002-0661-7517]{Ke Zhang}
\altaffiliation{NASA Hubble Fellow}
\affiliation{Department of Astronomy, University of Wisconsin-Madison, 
475 N Charter St, Madison, WI 53706, USA}
\affiliation{Department of Astronomy, University of Michigan, 323 West Hall, 1085 S. University Avenue, Ann Arbor, MI 48109, USA}

\author[0000-0003-1837-3772]{Romane Le Gal}
\affiliation{Center for Astrophysics \textbar\, Harvard \& Smithsonian, 60 Garden St., Cambridge, MA 02138, USA}
\affiliation{IRAP, Universit\'e de Toulouse, CNRS, CNES, UT3, Toulouse, France}
\affiliation{Univ. Grenoble Alpes, CNRS, IPAG, F-38000 Grenoble, France}
\affiliation{IRAM, 300 rue de la piscine, F-38406 Saint-Martin d'H\`{e}res, France}

\author[0000-0002-2692-7862]{Felipe Alarc\'on}
\affiliation{Department of Astronomy, University of Michigan, 323 West Hall, 1085 S. University Avenue, Ann Arbor, MI 48109, USA}

\author[0000-0003-1008-1142]{John~D.~Ilee} 
\affiliation{School of Physics and Astronomy, University of Leeds, Leeds, UK, LS2 9JT}

\author[0000-0003-1534-5186]{Richard Teague}
\affiliation{Center for Astrophysics \textbar\, Harvard \& Smithsonian, 60 Garden St., Cambridge, MA 02138, USA}

\author[0000-0003-2076-8001]{L. Ilsedore Cleeves}
\affiliation{Department of Astronomy, University of Virginia, Charlottesville, VA 22904, USA}

\author[0000-0003-1526-7587]{David J. Wilner}
\affiliation{Center for Astrophysics \textbar\, Harvard \& Smithsonian, 60 Garden St., Cambridge, MA 02138, USA}

\author[0000-0002-7607-719X]{Feng Long}
\affiliation{Center for Astrophysics \textbar\, Harvard \& Smithsonian, 60 Garden St., Cambridge, MA 02138, USA}

\author[0000-0002-6429-9457]{Kamber R. Schwarz}
\altaffiliation{NASA Hubble Fellowship Program Sagan Fellow}
\affiliation{Lunar and Planetary Laboratory, University of Arizona, 1629 E. University Blvd, Tucson, AZ 85721, USA}
 
\author[0000-0003-4001-3589]{Arthur D. Bosman}
\affiliation{Department of Astronomy, University of Michigan, 323 West Hall, 1085 S. University Avenue, Ann Arbor, MI 48109, USA}

\author[0000-0002-1199-9564]{Laura M. Pérez}
\affiliation{Departamento de Astronom\'ia, Universidad de Chile, Camino El Observatorio 1515, Las Condes, Santiago, Chile}

\author[0000-0002-1637-7393]{Fran\c cois M\'enard}\affiliation{Univ. Grenoble Alpes, CNRS, IPAG, F-38000 Grenoble, France}

\author[0000-0002-7616-666X]{Yao Liu}
\affiliation{Purple Mountain Observatory \& Key Laboratory for Radio Astronomy, Chinese Academy of Sciences, Nanjing 210023, China}




\begin{abstract}

Small organic molecules, such as \cch, HCN, and \hhco{}, are tracers of the C, N, and O budget in protoplanetary disks. We present high angular resolution ($10-50$~au) observations of \cch, HCN, and \hhco{} lines in five protoplanetary disks from the Molecules with ALMA at Planet-forming Scales (MAPS) ALMA Large Program. We derive column density and excitation temperature profiles for HCN and \cch{}, and find that the HCN emission arises in a temperate ($20 - 30$~K) layer in the disk, while \cch{} is present in relatively warmer ($20-60$~K) layers. In the case of HD~163296, we find a decrease in column density for HCN and \cch{} inside one of the dust gaps near $\sim83$~au, where a planet has been proposed to be located. We derive \hhco{} column density profiles assuming temperatures between 20 and 50~K, and find slightly higher column densities in the colder disks around T Tauri stars than around Herbig Ae stars. The \hhco{} column densities rise near the location of the CO snowline and/or millimeter dust edge, suggesting an efficient release of \hhco{} ices in the outer disk. Finally, we find that the inner $50$~au of these disks are rich in organic species, with abundances relative to water that are similar to cometary values. Comets could therefore deliver water and key organics to future planets in these disks, similar to what might have happened here on Earth. This paper is part of the MAPS special issue of the Astrophysical Journal Supplement.

\end{abstract}


\section{Introduction} \label{sec:intro}

Young stars have flattened disks of gas and dust rotating around them, a natural outcome of the star formation process. The material available in these protoplanetary disks will eventually be accreted onto the star, blown out by winds, or be incorporated into planets \citep[see reviews by][]{Williams2011,Andrews2020,Oberg2021}. The chemical composition of nascent planets is therefore directly linked to the chemical reservoir of disks. Small organics are some of the main carriers of C, N, and O in protoplanetary disks. However, the ability of planets to access these organic reservoirs depends on the distribution of this material, particularly in the inner few tens of au where rocky planets are expected to form. 

Although the main carrier of N in disks is expected to be N$_2$, this molecule has no observable lines in cold gas. HCN is a secondary N carrier, but its rotational lines in the millimeter are readily detected in disks \citep{oberg2011,chapillon2012,bergner2019}. In addition, observations of mid-infrared HCN lines have shown that it is abundant in the inner $\sim10$~au of disks \citep{salyk2011,najita2013,najita2018}. HCN is therefore one of our best tracers of the N budget in disks and allows us to characterize the spatial distribution of N. Moreover, HCN is thought to be a crucial molecule to the emergence of life, as it is the starting point in the chemistry leading to proteins and RNA \citep{powner2009,patel2015,becker2019}. 

In addition to revealing the distribution of the organic and elemental reservoir, molecular lines provide us key information on the physical and chemical conditions prevailing in disks. The distribution of HCN and \cch{} has been shown to depend on the UV radiation field and on the abundances of C and O \citep{du2015,bergin2016,cleeves2018}. The distribution of these small organics are also potentially sensitive to density and temperature variations. For example, HCN has been proposed to be a useful tool to study the formation of giant planets in disks, as they will heat up the surrounding gas and result in enhanced HCN emission around the forming planet \citep{cleeves2015}.

\hhco{} is a third commonly observed small organic in disks that is connected to the oxygen cycle. It is an important precursor of O-bearing Complex Organic Molecules (COMs), which as methanol, acetaldehyde,  ethylene glycol, and formamide \citep{watanabe2002,chuang2017}. It is hypothesised that O-bearing COMs are preferentially locked in ices and removed from the gas phase as the dust grows and settles to the midplane \citep[e.g,][]{bergin2010,Hogerheijde2011,du2017}. This would explain why O-bearing COMs have been more difficult to detect in disks compared to nitriles or hydrocarbons \citep[e.g.,][]{chapillon2012,oberg2015,walsh2016,bergner2018,Favre2018,loomis2018,carney2019}. \hhco{} lines, however, are much brighter in disks than lines from more complex O-bearing species because \hhco{} has a simpler molecular structure and therefore the internal energy is distributed in fewer molecular lines \citep{aikawa2003, pegues2020}. Observations of \hhco{} therefore provide a view of the spatial distribution of the O-bearing molecules in disks.


While rich substructure has been observed in dust continuum emission of numerous disks \citep[e.g.,][]{andrews2018,long2018,perez2020}, the chemical structure of these disks has been less explored at high angular resolution. The combination of sensitivity and spatial resolution needed to spatially resolve molecular line emission on planet-forming scales has been the main limitation so far in characterizing the distribution of most key organic molecules. However, sensitive observations at moderate angular resolution have demonstrated the presence of a rich substructure in the emission of various molecular lines  \citep[e.g.,][]{Henning2008,Mathews2013,oberg2015,bergin2016,huang2017,oberg2017,Salinas2017,Cazzoletti2018,Kastner2018,Miotello2019,pegues2020,teague2020}. 

In this paper, we present observations from the Molecules with ALMA at Planet-forming Scales (MAPS) ALMA Large Program. Here, we focus on molecular line emission from the small organic molecules HCN, \cch{}, and \hhco{} in five nearby and otherwise well-studied protoplanetary disks. HCN, \cch{}, and \hhco{} have been previously observed in these disks, but at lower angular resolution ($\sim$0\farcs5). In particular, \citet{bergner2019} presented observations of HCN and \cch{} toward four of these disks, and \citet{pegues2020} presented observations of \hhco{} toward the five disks. The MAPS data presented in this paper provide a combination of high angular resolution with excellent brightness sensitivity, which enables us to identify substructure in the line emission of these small organics. The observations are briefly described in Section~\ref{sec:obs}. In Section~\ref{sec:distribution}, we present the spatial distribution of the emission. In Section~\ref{sec:columndensity}, we retrieve the column density and excitation temperature profiles. We discuss the results in Section~\ref{sec:discussion} and summarize our conclusions in Section~\ref{sec:conclusons}.

\TabDiskSample{}
\TabSpecParams{}

\section{Observations} \label{sec:obs}

The observations are part of the MAPS ALMA Large Program (2018.1.01055.L), where a large number of molecular lines were observed at high spatial resolution to search for links between chemistry and planet formation in protoplanetary disks. See \citet{oberg21} for a detailed description of the scope and aims of MAPS. The MAPS sample includes disks around both T~Tauri (IM~Lup, GM~Aur, and AS~209) and Herbig Ae (HD~163296 and MWC~480) stars, with a wide variety of dust substructures (e.g., rings, gaps, and spirals). A detailed description of the sources can be found in \citet{oberg21}, and a brief summary is given in Table~\ref{tab:disk_sample}.

 Four spectral settings, two in Band 6 and two in Band 3, were observed between 2019 and 2020. The Band 6 setting covered the HCN $J=3-2$, \cch{} $N=3-2$, and the \hhco{} $3_{03} - 2_{02}$ lines. The Band 3 setting covered the HCN $J=1-0$ lines and the \cch{} $N=1-0$ ladder. See Table~\ref{tab:spec_params} for a full list of QNs or transitions observed. We subsequently just refer to these lines as $3-2$ and $1-0$ for simplicity. The data include observations using two different array configurations, to provide sufficient dynamic range in baseline length to recover both large-scale and small-scale emission. 

The data were first calibrated by the ALMA staff using standard routines, and then self-calibrated by the MAPS team to improve the signal-to-noise ratio of the data. A detailed description of the observations and calibration process can be found in \cite{oberg21}.

The imaging of the observed visibilities was done with the \texttt{tclean} task of the \texttt{CASA} 6 software. A Keplerian mask was created during the CLEANing process, by selecting regions in each channel where line emission is expected given the Keplerian rotation of each disk. Novel techniques were applied to improve the image quality, in particular the scaling of the CLEAN residual map, which can affect substantially the data quality when combining data from different configuration arrays. We encourage the reader to read a more detailed description of the imaging strategy in \citet{czekala21}.

We make use of the fiducial MAPS images, which have a circular beam of 0\farcs15 (corresponding to linear spatial scales of $15-24$~au) for the $3-2$ lines, and a circular beam of 0\farcs3 (corresponding to linear spatial scales of $30-48$~au) for the $1-0$ lines. In addition to these fiducial images, we used a second set of images for the $3-2$ lines, that were produced by applying a taper to match the resolution of the $1-0$ lines \citep[see Section 6.2 in][for a detailed description of how these images were produced]{czekala21}. This set of maps are available on the MAPS project homepage (\url{https://www.alma-maps.info}). The \hhco{} line is weaker per beam than that of the other small organics, resulting in images with low S/N at high angular (0\farcs15) resolution. GM~Aur is an exception here, as it presents bright \hhco{} emission that is well-detected in the 0\farcs15 resolution image. For consistency, here we use the tapered version of the \hhco{} images (with 0\farcs3 resolution) for all sources, which are more sensitive to low surface brightness and extended features. Disk integrated line fluxes for each line can be found in Table~\ref{tab:fluxes}.

\FigMaps{}

\section{Spatial distribution of the emission}
\label{sec:distribution}

\subsection{Zeroth-moment maps}

The velocity-integrated emission, or ``zeroth moment", maps of the HCN, \cch{}, and \hhco{} $3-2$ and $1-0$ lines are shown in Fig.~\ref{fig:maps}, the dust continuum emission at 260~GHz (with $\sim$0\farcs1 angular resolution) is shown in the upper row for comparison. A description of the continuum emission images can be found in \citet{sierra21}. For HCN $J=3-2$ and $J=1-0$, all hyperfine lines were included to generate the zeroth-moment maps, while for \cch{} only the two brightest (and blended) lines are included in the $N=3-2$ line, and the first and brightest component in the $N=1-0$ line. For more details on the zeroth-moment map generation process, see \citet{law21_rad}. 

\FigProfs{}

The $3-2$ lines have higher fluxes than the lower-energy $1-0$ lines for all disks. 
HCN $J=3-2$ is strongly detected in all disks, and presents a variety of substructures, including central holes, single, and multiple rings. HCN $J=1-0$ presents similar spatial distributions to the $J=3-2$ line in AS~209, HD~163296, and MWC~480. Several rings are seen toward HD~163296, including a faint ring in the outer disk (centered at $\sim400$~au), far beyond the dust continuum edge. Toward IM~Lup, the HCN $J=1-0$ line emission is fainter and more extended than in the other disks, resulting in a lower-S/N zeroth-moment map. Toward GM~Aur, the HCN $J=1-0$ line is mostly detected in the central regions of the disk, although an outer diffuse ring is seen in the radial profile \citep[see also][]{law21_rad}. 

The \cch{} $N=3-2$ and $N=1-0$ lines have spatial distributions and radial extents similar to HCN in AS~209, HD~163296, and MWC~480. Toward IM~Lup and GM~Aur, the $N=3-2$ line is fainter than in the other disks, resulting in lower-S/N zeroth-moment maps, and the $N=1-0$ line is mostly detected in the central regions of the disks, similar to HCN. 

Fewer substructures are observed for \hhco{} compared to HCN and \cch{}, although a few ring-like structures are seen, in particular in HD~163296. The \hhco{} line emission is more extended than HCN and \cch{}, and presents a central hole of varying depths in all the disks except GM~Aur and MWC~480, where only a tentative dip is observed. GM~Aur presents exceptionally bright emission compared to the other disks. 

Normally, for molecules other than \twco{} and \thco{}, it is not possible to derive the height of the emission directly from the spatially resolved image cubes \citep[e.g.,][]{rosenfeld2013,pinte2018b}. This is because other lines are typically fainter and arise in less elevated disk layers. However, thanks to the higher inclination and relative proximity of the HD~163296 disk, and the high angular resolution and sensitivity of the MAPS data, we can distinguish emission arising in a flat disk from emission arising in the elevated layers above the disk midplane. Indeed, the HCN and \cch{} $3-2$ lines observed toward HD~163296 show a clear asymmetry along the disk minor axis that cannot be explained with emission from a flat disk. Part of the asymmetry, also seen as projected ellipses that look shifted to the south with respect to the disk center, is due to the fact that we are observing the emission from the near cone that is facing us. This effect is clearly seen in the zeroth-moment map, and corresponds to the heights estimated in \citet{law21_surf}, i.e., $z/r \lesssim 0.1$. See \citet{law21_surf} for a detailed description of the emission surfaces for HCN and \cch{} in the MAPS disk sample.

\subsection{Radial profiles}

To better characterize the spatial distribution of the emission we used the \texttt{radial\_profile} function of the Python package \texttt{GoFish} \citep{Teague2019} to produce deprojected radial intensity profiles from the zeroth-moment maps, considering the disk inclination and position angle (see Table~\ref{tab:disk_sample}). Although it is possible to take into account the vertical height of the emission to derive the radial intensity profiles \citep{law21_rad}, here we assume a flat disk but produce the deprojected profiles by averaging the emission within a $30^{\circ}$ wedge along the disk major axis. The \cch{} lines were not detected at sufficiently high S/N toward IM~Lup and GM~Aur, so we averaged the emission azimuthally to produce the deprojected radial profiles. For additional details on the deprojected radial profiles, see \citet{law21_rad}. The resulting profiles for the $3-2$ (solid) and $1-0$ (dashed) lines are shown in Fig.~\ref{fig:profs}. For comparison, the profiles of the 260~GHz and 90~GHz dust continuum emission are shown in the first row.

In general, the HCN and \cch{} emission is more compact than that of \hhco{} --- their radial extents are smaller by a factor $\sim2$. Moreover, their radial extents are similar to that of the dust continuum emission, although faint HCN and \cch{} emission is observed up to $\sim400$~au in HD~163296. IM~Lup is an exception, as it presents similar radial extents in all molecular lines. One possibility to explain the lack of HCN and \cch{} emission in the outermost disk regions where \hhco{} is observed is that the \hh{} gas density is not high enough to excite the lines. Indeed, for gas temperatures between 10 and 50~K, the HCN $3-2$ line has a critical density of $(6-10)\times10^6  \pccm$, while the \hhco{} $3-2$ line has a critical density that is 10 times lower \citep{Shirley2015}. However, models of these disks find that gas densities are $>10^7 \pccm$ in the outer disk midplanes, enough to excite these lines \citep[e.g.,][]{zhang21}. This suggests that the abundance of HCN and \cch{} decreases in the outermost disk regions where \hhco{} is still abundant. We note that no additional emission arises in the outermost disk regions in the lower-resolution (0\farcs3) images that have better S/N.

The radial profiles are consistent with profiles previously observed at lower angular resolution (0\farcs5) for HCN and \cch{} in IM~Lup, AS~209, HD~163296, and MWC~480 \citep{bergner2019}, and for \hhco{} in all sources \citep{pegues2020}. However, our higher angular resolution observations reveal additional radial substructures, in particular for HCN and \cch. We focus on the $3-2$ lines which show more structure than the $1-0$ lines, due to the higher signal-to-noise ratios and angular resolution \citep[see][for additional substructure analysis]{law21_rad}. 

\subsubsection{HCN}

HCN shows a central hole in four of five disks, the exception being MWC~480 where the emission is centrally peaked. A main ring of emission is seen for IM~Lup, GM~Aur, and AS~209, although the rings are not Gaussian but present one or more emission shoulders that could result from blended ring components not resolved in the MAPS data. IM~Lup shows an unusually broad, almost plateau-like ring. Two bright rings are seen for HD~163296, centered at $\sim30$ and $110$~au, and a fainter plateau that extends up to $\sim400$~au. In addition, a shallow gap is seen at $\sim327$~au along with a faint ring centered at $\sim366$~au \citep{law21_rad}. The ring is better seen in the zeroth-moment map (see Fig.\ref{fig:maps}). GM~Aur and AS~209 also present faint HCN emission in the outer disk. In addition to the central emission component, MWC~480 shows an emission shoulder at $\sim30$~au, and a partially blended ring at $\sim90$~au \citep[see also][]{law21_rad}. HCN radial profiles observed at similar linear spatial resolution toward other disks around T~Tauri stars generally show centrally peaked emission profiles with emission bumps in the outer disk \citep[e.g.,][]{hily-blant2019,Kastner2018}.

\subsubsection{\cch{}}

\cch{} shows a central hole for all disks, which is generally larger than the hole seen for HCN. Several unresolved ring components are seen toward IM~Lup and GM~Aur, similar to what is observed in HCN. AS~209 presents a single ring of \cch{} emission, very similar in size to the ring seen in HCN. Two rings are seen toward HD~163296, centered roughly at the same locations of the HCN rings. The two faint HCN rings seen in the outer disk are also observed for \cch{} in the zeroth-moment map (see Fig.~\ref{fig:maps}), but are harder to see in the radial profiles. The presence of these faint \cch{} outer rings is confirmed with the tapered (0\farcs3 resolution) image, which has higher S/N \citep{law21_rad}. Toward MWC~480, a single \cch{} ring is observed centered at the same location of the outermost HCN ring. Interestingly, the innermost \cch{} ring seen in HD~163296 and the rings seen in AS~209 and MWC~480 are all located near a gap in the dust continuum emission. Similar ring-like structures with large central cavities have been observed for \cch{} toward other disks, like TW~Hya \citep{bergin2016} and V4046~Sgr \citep{Kastner2018}.

\subsubsection{\hhco{}}
\label{sec:prof_h2co}

\hhco{} radial profiles present more complicated substructure, such as broad ring-like emission components with several shoulders and/or local maxima. In the inner disk, IM~Lup, AS~209, and HD~163296 present a central depression in \hhco{} emission, while GM~Aur and MWC~480 show centrally peaked emission profiles. In the outer disks ($>100$~au), the slope at which the \hhco{} intensity decreases is much more shallow compared to HCN and \cch{}, in particular for the disks around HD~163296 and MWC~480. 

The radial profiles show spatial links between \hhco{} substructures, namely line emission peaks, and the outer edge of the millimeter continuum disk. \hhco{} peaks are aligned with the outer edge of the continuum in IM~Lup and HD~163296. Additionally, the higher-resolution (0\farcs15) images of AS~209, as noted in \cite{law21_rad}, also show the presence of an additional \hhco{} ring that is radially coincident with the dust edge. Toward HD~163296, a spatial link between a \hhco{} line emission peak and outer dust edge was seen in lower-resolution data from \citet{carney2017}. Spatial links between millimeter dust edges and \hhco{} line emission have also been observed toward other disks, like TW~Hya \citep{oberg2017}, and V4046~Sgr \citep{Kastner2018}, as well as in four out of a sample of 15 disks in the survey presented by \citet{pegues2020}. All MAPS disks were included in those surveyed by \cite{pegues2020}, but a close association between an outer \hhco{} ring and millimeter dust edge in the MAPS disks, as reported here, was not clearly seen in \cite{pegues2020}, due to these outer \hhco{} rings not being clearly spatially resolved. Thus, when considering these previous survey results with the new MAPS observations, this suggests that at least ${\sim}$50\% of disks may show some spatial links between \hhco{} substructures and the outer edge of the continuum disk. 
\\

In summary, we find some similarities between the HCN and \cch{} radial profiles. Furthermore, several of the \hhco{} peaks spatially coincide with the HCN and \cch{} peaks, for example the first \hhco{} peak seen toward AS~209. However, most of the \hhco{} features do not have direct counterparts in HCN or \cch{}, and the \hhco{} radial extents are larger.

\section{Column densities} \label{sec:columndensity}

To derive column density profiles, we first produce deprojected spectra at different radii using the \texttt{radial\_spectra} function of \texttt{GoFish}. The radial bins are set to 1/4 of the beam major axis, and the disk is assumed to be vertically flat. Before averaging the spectra in each radial bin, the velocity is shifted according to the Keplerian rotation of the disk, so that all the spectra are aligned. The stellar masses needed to compute the Keplerian velocities are taken from \cite{oberg21} and are estimated from gas dynamics in \citet{Teague21}. The spectra are azimuthally averaged in those disks with low S/N (\cch{} for IM~Lup and GM~Aur, and \hhco{} for all sources), and averaged within a $30^{\circ}$ wedge along the disk major axis for the others. The main motivation for using a $30^{\circ}$ wedge when possible is to avoid the uncertainty of the emission surface height, which strongly influences the spatial location of emission near the minor axis. 

\subsection{Column density retrieval} 

To derive the column densities, we assume that the emission is in local thermal equilibrium (LTE) and that the emission fills the beam. The LTE assumption is reasonable, considering the high densities present in protoplanetary disks. \citet{Cataldi21} performed a non-LTE fit to the HCN lines and found similar column densities, suggesting our LTE assumption is a good approximation, at least for HCN. The line intensity is computed as
\begin{equation}
    I_{\nu} = (B_{\nu}(\Tex) - B_{\nu}(T_{bg})) [1 - \exp(-\tau_{\nu})]
\end{equation}
where $B_\nu$ is the Planck function, $\Tex$ is the excitation temperature, $\Tbg$ is the background temperature, and $\tau_\nu$ is the line optical depth. The background temperature $\Tbg$ is fixed to the maximum of the continuum brightness temperature or the cosmic microwave background (CMB) temperature (2.73~K).

At the line center, the optical depth is given by
\begin{equation}
    \tau_0= \frac{ A_{ul} c^3 N_u}{ 8 \pi \nu^3 \sigma \sqrt{2\pi} } \left [ \exp \left (\frac{h \nu}{k \Tex} \right ) -1 \right ] 
\end{equation}
where $A_{ul}$ is the Einstein coefficient, $\nu$ is the frequency of the line, $\sigma$ is the Gaussian line width and N$_u$ is the upper-level column density,
\begin{equation}
     N_u = \frac{N_{\rm{tot}}}{Q(\Tex)} g_u \exp \left ( \frac{-E_u}{k \Tex} \right ).
\end{equation}
Here, N$_{\rm{tot}}$ corresponds to the total column density of the molecule, $g_u$ is the upper-level degeneracy, $E_u$ is the upper-level energy, and $Q(\Tex)$ is the partition function. The spectroscopic constants and partition functions were taken from the CDMS and JPL catalogs and are listed in Table~\ref{tab:spec_params}.

The free parameters in the fit are the total column density N$_{\rm{tot}}$, the FWHM $\Delta V = 2 \sqrt{2 \ln2} \sigma$, and the excitation temperature in the case of HCN and \cch{}, as their lines present hyperfine structure. For \hhco{}, the excitation temperature is fixed because only a single line is available. We also allow for a small offset from the source velocity $v_{lsr}$, constrained between $-$0.25 and 0.25$\kms$. We explore the parameter space using the \texttt{emcee} package \citep{Foreman2013}. The best fit is taken as the 50th percentile of the samples in the marginalized distributions, and the associated errors are computed from the 16th and 84th percentiles.

\subsection{HCN and \cch{} fitting method}

\FigSpecFitHighRes{}

We follow a method similar to the one implemented in \citet{bergner2019}. The HCN and \cch{} lines present hyperfine structure that is spectrally resolved. We fit the observed spectra assuming that all the hyperfine lines have the same excitation temperature. Each hyperfine line is assumed to be a Gaussian, such that the total line opacity is given by
\begin{equation}
    \tau_{\nu} = \sum_i \tau_0 r_i \exp \left ( -\frac{(v - v_i - v_{lsr})^2}{ 2 \sigma^2} \right) 
\end{equation}
where $\tau_{i,0}$ is the line center opacity of the {\it i}th component, $v_i$ is the relative velocity of each component with respect to the brightest line, and $v_{lsr}$ is the source velocity. The relative intensities $r_i$ as well as the relative velocities $v_i$ are listed in Table~\ref{tab:spec_params}. 

The width of spectral lines is dominated by thermal broadening, which depends on the gas temperature:
\begin{equation}
    \sigma_{\rm therm} = \sqrt{ \frac{k \Tgas}{m} },
\end{equation}
where $m$ is the molecule's mass. Under the assumption of LTE, $\Tgas=\Tex$. However, we find that the observed line widths for HCN and \cch{} are broader than their thermal widths would imply. This is most likely produced by beam smearing, where line emission from different radii in the disk that have different velocities are combined in a single beam. In addition to beam smearing, it is possible that emission from both the back and front sides of the flared disk, which have slightly different projected velocities, are being combined in the observed spectra. The inner disk is particularly affected by these issues, which makes it difficult to fit the lines without artificially decreasing the line opacity or increasing the excitation temperature. A way to alleviate this problem is to first compute the line intensity assuming a purely thermal line width, then compute the total flux, and finally redistribute this flux in the line but now for the observed and broader line width. With this method, the fit to the lines is better, and the inferred line opacity---and hence the column density---is more realistic. The same solution was applied to fit the CN and DCN lines in the MAPS disks \citep{bergner21,Cataldi21}.

The fit to the observed spectra was done in two steps. First, we fit the $1-0$ and $3-2$ lines simultaneously to obtain a better constraint on the excitation temperature. The tapered (0\farcs3) versions of the $3-2$ line cubes were used for this step, to match the resolution of the $1-0$ line cubes. We included priors on the excitation temperature informed by the gas temperature traced by the optically thick \twco{} 2--1 line. \citet{law21_surf} found that the HCN and \cch{} emission arises from relatively flat surfaces $z/r\lesssim0.1$ in AS~209, HD~163296, and MWC~480, while the CO isotopologues arise in more elevated disk layers. We thus restrict the rotational temperature between 10~K and the gas temperature inferred from the peak brightness temperature of the optically thick \twco{} 2--1 line, $\Tgas$(\twco) \citep{law21_surf}, for both HCN and \cch{}. The results for the simultaneous fit to the $1-0$ and $3-2$ lines are shown in Appendix~\ref{app:fit}. The resulting excitation temperature and column density profiles are shown in Fig.~\ref{fig:NprofsCombined}, and the gray curves show the $\Tgas$(\twco) profiles for each disk. HCN presents rotational temperatures between 15 and 40~K, while \cch{} presents higher values of $15-60$~K. The column densities are on the order of $10^{13} - 10^{15} \pscm$ for both molecules. Despite the lower angular resolution of 0\farcs3 of the data used in these analyses, two column density rings are resolved in both HCN and \cch{} for HD~163296, and a single ring or central component for the other sources. 

In the second step, and to obtain a better-resolved column density profile, we fit the high angular resolution $3-2$ lines only. We used the excitation temperature found in the previous step (T$_{\rm{ex},0}$) as an initial parameter, and constrain $\TexO - 10$~K $< \Tex <  \TexO + 10$~K for both HCN and \cch{}. Figure~\ref{fig:SpecFitB6} shows a selection of observed spectra and fit for each species and disk. Note that the radial extent is different for each disk. The lines are fit well in most of the disk, but poorly fit in the inner disk, where the lines are substantially broader, resulting in the individual hyperfine components being completely blended. For this reason, we used the \twco{} or \thco{} temperature as the excitation temperature in regions where the fit does not work properly. For \cch{}, the $\Tgas$(\twco) was used for all sources. For HCN, the $\Tgas$(\twco) was used for IM~Lup, GM~Aur, and AS~209. For the two disks around Herbig Ae stars, however, the $\Tgas$(\thco) was used instead, since the $\Tgas$(\twco) produced a large discontinuity in the column density profile. 

An independent retrieval of the HCN column density and excitation temperature is presented in \citet{bergner21} and \citet{Cataldi21}, using a method similar to the one presented here but with different constraints on the excitation temperature. A comparison between the HCN column densities presented in these three papers is shown in  Appendix~\ref{app:HCNcomparison}. The resulting column density profiles and excitation temperatures are consistent with the ones presented here (see Fig.~\ref{fig:Nprofs-compHCN}).

\FigNprofs{}

\subsection{\hhco{} fitting method}

For \hhco{}, we fit a single Gaussian to the $J=3-2$ line in each radial bin. In order to constrain the \hhco{} excitation temperature, a combination of lines with different upper level energies is needed. Since we only observed one line, the excitation temperature is fixed and the column density was computed for a range of excitation temperatures between 20 and 50~K so that we are not {\it a priori} assuming the vertical location of the \hhco{} emitting layer. The range of temperatures is based on previous estimates. For example, \citet{pegues2020} recently showed that the bulk \hhco{} emission in four Class II disks arises in $11 - 40$~K gas, using multi-line \hhco{} observations with similar energy ranges ($E_u=20-70$~K). In particular, for MWC~480 they found a low excitation temperature of $\sim21$~K.  Similarly, toward HD~163296 the disk-averaged \hhco{} excitation temperature has been estimated to be $\sim 24$~K \citep{carney2017,guzman2018}. Finally, toward the older \citep[$\sim8$~Myr;][]{Sokal2018} TW~Hya disk, \citet{Terwisscha2021} recently found that \hhco{} emits mostly from a $30-40$~K layer, corresponding to an elevated disk layer of $z/r \geq 0.25$. 

We found that the column density is not very sensitive to $\Tex$; the difference in column density is only a factor $\sim2$ for this range of temperatures. The observed spectra with the best-fit overlaid are shown in the bottom panels in Fig.~\ref{fig:SpecFitB6}. As was found for HCN and \cch{}, the lines are fit well in most of the disk, but poorly fit in the inner $\sim30$~au, especially in HD~163296 and MWC~480. This is probably due to contamination from the back side of the disk. 

\subsection{HCN, \cch{}, and \hhco{} column density profiles}

The resulting column density profiles for HCN, \cch{}, and \hhco{} are shown in Fig.~\ref{fig:Nprofs}. Two profiles are shown for \hhco{}, corresponding to $\Tex$ of 20 and 50~K. The gray shaded area marks the inner regions where $\Tgas$(\twco) or $\Tgas$(\thco) was used as the excitation temperature in the fit. The horizontal bars show the spatial resolution of the profiles for each molecule and disk.  


The HCN and \cch{} column densities range from $10^{12}$ to $10^{15} \pscm$, although the HCN column density may reach $\gtrsim10^{16} \pscm$ in the inner $30$~au of GM~Aur, HD~163296, and MWC~480. Overall, HCN and \cch{} present similar column density profiles, except for that HCN has a central component in GM~Aur, HD~163296, and MWC~480, while the inner disk seems to be depleted of \cch{} in all sources but toward GM~Aur. GM~Aur presents a central dust cavity (see Fig.~\ref{fig:profs}), which could result in enhanced \cch{} abundance in the inner disk. However, there is evidence that the total \hh{} gas is also depleted inside the dust cavity \citep{Dutrey2008,Huang2020,bosman21b}. Another difference between HCN and \cch{} is in the outer $>150$~au disk. The HCN column density seems to reach a plateau of a few times $10^{12} \pscm$, which is not present for \cch{}.

\FigTexprofs{}

The \hhco{} column density profiles are consistent with the ones derived by \citet{pegues2020} for all sources using different \hhco{} transitions and at lower angular resolution. \hhco{} is present at higher column density at larger disk radii compared to HCN and \cch{}. Although the spatial resolution for the \hhco{} column density profiles is lower than for the smaller organics, a few distinct rings can be seen in the profiles. GM~Aur and AS~209 present less structure, while at least two rings or local maxima are seen for HD~163296 and MWC~480. The \hhco{} column density varies between the sources, ranging between $10^{12}$ and $10^{14} \pscm$ in the inner 200~au. The largest column densities are seen toward GM~Aur and IM~Lup, which together with AS~209 are the coldest disks in the sample if we consider their stellar mass and the large disk masses that have been estimated \citep{cleeves2016b,McClure2016}. 

\subsection{HCN and \cch{} excitation temperatures}

Figure~\ref{fig:Texprofs} shows the excitation temperature profiles inferred for HCN and \cch{} from fitting the higher angular resolution $3-2$ lines only, informed by the $\Tex$ obtained from the simultaneous fit to the $1-0$ and $3-2$ lines. The $\Tgas$(\twco) and $\Tgas$(\thco) profiles are shown in gray. At least for 3/5 sources, the \cch{} excitation temperatures are higher than those derived for HCN. HCN temperatures range from 20 to 30~K, while \cch{} temperatures range from 20 to 60~K. The \cch{} excitation temperatures in the outer disk regions ($>100$~au) of AS~209, HD~163296, and MWC~480 are consistent with the temperatures derived by \citet{bergner2019} at lower angular resolution. Excitation temperatures for HCN do not vary substantially within the disk sample. However, the inferred $\Tex$ values for \cch{} are higher for HD~163296 and MWC~480 than for the other three disks. This is consistent with disks around Herbig~Ae stars being warmer than disks around lower-mass T~Tauri stars, and suggest that \cch{} arises in warmer elevated disk layers compared to HCN.

The inferred excitation temperatures are low ($20-30$~K) for both HCN and \cch{} in IM~Lup and GM~Aur, and do not vary substantially across the disks. Toward AS~209, the \cch{} excitation temperature profile shows a ring-like distribution similar to the column density ring seen in Fig.~\ref{fig:Nprofs}, reaching a peak $\Tex\sim40$~K in one of the dust gaps. HCN, on the other hand, shows a decreasing $\Tex$ profile with radius. Toward HD~163296, the \cch{} and HCN excitation temperatures show two rings, similar to the distribution of the column densities, although the rings are more pronounced for \cch{} than for HCN.  Toward MWC~480, the $\Tex$ of HCN and \cch{} decrease with radius in a smoother profile compared to HD~163296. Finally, in AS~209 and HD~163296, the $\Tex$ peaks of \cch{} spatially coincide with the column density peaks. We do not find the same correlation for the other sources or for HCN. 

\section{Discussion} \label{sec:discussion}

\FigNprofsComparisonCO{}

\subsection{Location of the HCN and \cch{} emission surfaces}

Chemical models predict HCN and \cch{} emission arises in elevated disk layers where UV photons drive the gas to temperatures of $40-50$~K and dominate the chemistry \citep{bergin2016,cleeves2018}. However, a recent analysis of observations toward a sample of 16 disks at lower ($>$0\farcs5) angular resolution resulted in low excitation temperatures of $5-10$~K and $10-40$~K for HCN and \cch{}, respectively \citep{bergner2019}.  \citet{bergner2019} suggested that the origin of the lower-than-expected excitation temperatures could be due to subthermal emission or beam dilution due to unresolved chemical substructure. The higher angular resolution MAPS data reveals these substructures across a sample of disks for the first time. In addition, the $1-0$ line observations now allow us to better constrain the molecular excitation temperature. While we do find HCN excitation temperatures that are higher than inferred by \citet{bergner2019} toward these disks, they remain relatively low ($<30$~K). For \cch{}, on the other hand, we find higher excitation temperatures that are more consistent with the gas temperature at the expected height of the emission, in particular for the warmer disks HD~163296 and MWC~480, which are also disks that present the most substructure in their emission. 

The inferred excitation temperatures for HD~163296 are consistent with the brightness temperature distributions of the HCN and \cch{} $3-2$ lines found by \citet{law21_surf}. These brightness temperatures provide a lower limit to the gas temperature because these lines are partly optically thin. These results, combined with the emission heights estimated by \citet{law21_surf} of $z/r\lesssim0.1$, suggest that the HCN emission arises in vertically deeper and colder layers than \cch{}. 

For IM~Lup and GM~Aur, the $\Tex$ of \cch{} and HCN are very similar, which suggests that their emission arises from similar disk layers. However, given our uncertainties, we cannot rule out a vertical stratification. Indeed, detailed chemical models of IM~Lup suggest that HCN arises in deeper layers compared to \cch{}. \citep{cleeves2018}. Finally, it is worth mentioning that the \cch{} excitation temperature is lower for the colder disks around T~Tauri stars, which show fewer substructures compared to the disks around Herbig~Ae stars. This could be due to intrinsic lower gas temperatures or to small-scale substructure not resolved in our observations. 

\subsection{Comparison to CO column densities}

CO is widely used to trace the gas in disks because \hh{} does not have strong transitions in the low temperature range that characterizes disks. Thus, we compare the column density profiles of the small organics with that of CO. Figure~\ref{fig:Nprofs-comp} shows the normalized column density profiles of HCN, \cch{}, and \hhco{}, as well as the CO column densities derived from the C$^{18}$O 2--1 line by \citet{zhang21}. In general, the HCN and CO column densities tend to be more centrally peaked than that of \cch{}, suggestive of an active warm cyanide chemistry in the inner disk. In the outer disk, HCN and \cch{} show similar radial distributions up to $\sim150$~au, where the \cch{} column density starts to drop faster than HCN and CO. There is notably good spatial correlation between the profiles toward HD~163296 between $50<r<150$~au. 

In general, the HCN and CO column densities correlate well, except in AS~209, where significant differences are observed. In particular, an anticorrelation is observed between CO and both HCN and \cch{}. \citet{Alarcon21} used chemical models to investigate the gas structure in AS~209, and showed that this anticorrelation can be explained by CO chemical processing, which produces a decrease in the CO abundance and an elevated C/O ratio ($>2$) in the gas that boosts the formation of \cch{}. \citet{bosman21a} also found that elevated C/O ratios ($\sim2$) are needed to explain the observed \cch{} column densities for AS~209, MWC~480, and HD~163296. 
\cch{} is therefore very sensitive to the gas-phase C/O abundance ratio in the disk atmosphere \citep{bergin2016,cleeves2018,Miotello2019}. 

\subsection{Photochemistry of HCN and \cch{}}

If the formation of HCN and \cch{} is the result of a UV-dominated chemistry, we might expect to see enhanced column densities inside the dust gaps. The depletion of millimeter dust at these locations should allow the deeper penetration of UV photons, and thus efficient formation of \cch{} and HCN. However, the penetration of UV photons will also depend on the geometry of the source (for example, the flaring of the disk). It is nevertheless worth comparing the structures observed in mm dust and molecular emission. The position of the dust continuum gaps, as measured by \citet[][; but see references therein]{law21_rad}, are marked by the gray dashed vertical lines in Fig.~\ref{fig:Nprofs-comp}. We only include the gaps that are resolved at $\sim0\farcs1$ resolution. However, several other gaps are seen in higher angular resolution observations of the dust continuum emission \citep{huang2018,Huang2020}. The peak of the \cch{} column density seen toward IM~Lup, AS~209, MWC~480, and the first peak seen toward HD~163296, coincides with the location of the innermost dust gaps, although the \cch{} structures are generally broad compared to the dust gaps. The second column density peak seen at $\sim110$~au toward HD~163296 does not spatially coincide with any of the dust gaps, revealing that there is not a universal connection between dust depletion and \cch{} or HCN formation. The weak correlation between dust structures and HCN and \cch{} chemical substructures could be due to the HCN and \cch{} emission arising in disk elevated layers, which are not always impacted by the dust millimeter substructure that trace the midplane. Moreover, recent studies have shown that the \hh{} density seems to be much less altered than the dust inside the dust gaps, and chemical effects play a more important role in setting molecular abundances \citep[e.g.,][]{Teague2018a,Alarcon21,zhang21}. 

If photochemistry drives the formation of HCN and \cch{}, we could also expect to see enhanced emission in the outer disk where the gas density starts to drop allowing UV photons to penetrate the disk. The flaring of the outer disk could also help to capture more UV photons from the star. Although the HCN and \cch{} emission is compact compared to \hhco{}, some faint emission is indeed detected at large radii, far beyond the dust millimeter edge in IM~Lup, GM~Aur, and HD~163296. As an example, Fig.~\ref{fig:ProfHD} shows radial structures seen in the tapered images for HD~163296, the disk that presents the most substructure in molecular line emission. A power-law stretch was used in the y-axis, to enhance the low-S/N structure seen in the outer disk. A faint ring that is seen in both HCN and \cch{} near 400 au is also seen in CN \citep{bergner21}. The ring is not clearly seen in \hhco{} in the radial profile, but a ring structure is visible at large radii in the zeroth-moment map, in particular along the disk minor axis (see Fig.~\ref{fig:maps}). No millimeter dust grains are present at these larger radii, but small dust grains are present, as suggested by scattered light images \citep[e.g.,][]{monnier2017,Avenhaus2018,rich2020}. A small dust grain population is also present in the disk elevated layers where HCN and \cch{} emission seems to arise. This small grain population is likely regulating the HCN and \cch{} chemistry. An increased amount of UV photons will also contribute to enhanced HCN destruction in the outer disk. Indeed, \citet{bergner21} find an increasing CN/HCN profile with radius, indicative of increased HCN photodissociation relative to CN in the lower-density outer disk \citep[see also][]{guzman2015}. All this suggests that an active photochemistry is at play in the outer disk beyond the millimeter dust emission around HD~163296. Even so, as mentioned in the previous section, the C/O ratio in the gas may still be more important than UV exposure for the formation of HCN and \cch{} in the outer disk. 

\subsection{Links to planet formation}

\FigProfsHD{}

Several efforts have been made in recent years to find planets forming in protoplanetary disks. The presence of several planets has been proposed to explain the gaps seen in the dust continuum emission and the deviation from Keplerian rotation in CO observed in a few disks, including HD~163296 \citep{teague2018b,pinte2018,pinte2020,Teague21} and MWC~480 \citep{Teague21}. 
While dust depletion in the midplane can be inferred from the millimeter dust continuum observations, it has been harder to infer the gaps are also depleted in gas, since all millimeter lines, including CO isotopologues, may reflect chemistry variations rather than true gas density variations. Indeed, chemistry has a nonlinear response to gas density. See \citet{rab2020} for detailed thermochemical modeling of HD~163296 in particular. 
Here, we compare the locations of the planets proposed for HD~163296 and MWC~480 with the structures observed in HCN, \cch{}, and \hhco{}.

In Fig.~\ref{fig:ProfHD}, we compare the emission of HCN, \cch{}, and \hhco{} with the locations of the putative planets in HD~163296 (dashed vertical lines). Besides the two bright rings seen in the inner 150~au, two fainter rings are resolved in the outer disk for both HCN and \cch{} \citep[see also][]{law21_rad}. The planet predicted at 83~au by \citet{teague2018b} and \citet{Isella2016} is located inside a dust millimeter gap, which coincides with the gap seen in HCN and \cch{}. This gap is also recovered in the column density profiles (Fig.~\ref{fig:Nprofs}). The correlation between the planet location and a decrease in the \cch{} column density was previously hinted by \citet{bergner2019}. Although the gaps seen for HCN and \cch{} are much shallower than what is observed for the millimeter dust, these observations suggest the gap is also depleted of gas.

Toward MWC~480, \citet{Teague21} proposed that a planet near 245~au could explain the observed CO deviations from Keplerian rotation. However, HCN and \cch{} are not detected at these large radii. At smaller radii ($<200$~au), no perturbations could be confirmed in the CO gas kinematics. However, a gap is seen in HCN at around 70~au, which roughly corresponds to the location of a millimeter dust gap. The gap is not seen in \cch{}. Hydrodynamical simulations suggest that a $\sim2.3$~M$_{\rm J}$ planet located at an orbital radius of $\sim78$~au could produce the observed dust gap \citep{liu2019}. If there is a corresponding gas gap, then this suggests that HCN may be more sensitive to gas density perturbations than \cch{}, and could potentially be used as a diagnostic to identify massive planets in protoplanetary disks.



\TabOrganicreservoir{}

\subsection{Origin of the \hhco{} emission}

\hhco{} could be a better tracer of the colder gas in the outer disk than HCN and \cch{}, which seem to trace warmer elevated layers in the disks. The use of \hhco{} as a tracer of cold gas depends on its dominant formation pathway and on the efficiency of desorption mechanisms. \hhco{} can form from both gas phase and grain surface chemistry. The dominant reaction in the gas-phase at low ($\lesssim100$~K) temperatures is the neutral-neutral reaction 
\begin{equation}
\rm{CH_3 \xrightarrow{O} H_2CO} 
\end{equation}
The grain surface chemistry pathway involves the successive hydrogenation of CO ices \citep{hidaka2004,watanabe2004,fuchs2009}
\begin{equation}
\rm{CO \xrightarrow[]{H} HCO \xrightarrow[]{H} H_2CO \xrightarrow[]{H} H_3CO \xrightarrow[]{H} CH_3OH. }    
\end{equation}
This pathway can only be efficient in regions where CO has started to freeze out onto dust grains, i.e., beyond the CO snowline and below and around the CO snow surface, depending on whether the CO hydrogenation is faster than the CO desorption. The light blue vertical lines in the bottom panels of Fig.~\ref{fig:Nprofs-comp} mark the estimated location of the CO snowline in each disk as measured by \citet{zhang21}. CO snowlines are located at smaller radii in the colder disks around T~Tauri stars, compared to the warmer HD~163296 and MWC~480 disks, where the CO snowline is well-resolved by our observations. The first \hhco{} ring seen toward HD~163296 and MWC~480 spatially coincides with the estimated location of the CO snowline, suggesting that CO hydrogenation is an important (if not dominant) formation pathway of \hhco{} in disks. 

In the warm inner disk, \hhco{} is expected to form mainly in the gas phase, since CO cannot freeze out onto dust grains. The inner regions of the MAPS disks all show central \hhco{} depressions, although they are less pronounced for GM~Aur and AS~209. This does not mean that \hhco{} is depleted in the inner disk; it could purely be a result of the observed line being more sensitive to cold gas. Indeed, higher-energy lines have shown centrally peaked emission profiles suggesting a warmer \hhco{} component is present in the inner disk \citep{loomis2015,oberg2017}. 

In the outer disk, both gas-phase and CO hydrogenation could contribute to the overall \hhco{} emission. In Section~\ref{sec:prof_h2co} we showed that at least ${\sim}$50\% of disks may show some spatial links between \hhco{} substructures and the outer edge of the continuum disk. We also observe spatial links between the \hhco{} column density profiles and the outer edge of the continuum disk, which are marked by the gray vertical lines in Fig.~\ref{fig:Nprofs-comp}. In particular, the second ring or bump component seen in the \hhco{} column density profiles of HD~163296 and MWC~480 spatially coincides with the millimeter dust edge. A similar coincidence is seen toward IM~Lup and GM~Aur, where a shoulder-type structure is seen near the continuum edge. This suggests that the release of ices at large distances becomes more efficient, either through thermal desorption due to thermal inversion \citep{cleeves2016}, or by nonthermal desorption due to the penetration of UV photons \citep[e.g.,][]{podio2020,vanthoff2020}. Another possibility is that other ice fragments are released into the gas phase in the outer disk, which then boost the gas-phase chemistry enhancing the abundance of \hhco{}. 

The \hhco{} traced by our observations could also form predominantly in the gas phase through reaction (6), with little contribution from the CO hydrogenation pathway (7). This could happen if nonthermal desorption of \hhco{} ices is not efficient and \hhco{} remains frozen onto dust grains. In this scenario, the emission would arise in more elevated and warmer disk layers. Although chemical models have shown that pure gas-phase chemistry cannot account for the observed \hhco{} abundances in the outer disk around DM~Tau \citep{loomis2015}, observations of edge-on Class I disks have shown that \hhco{} emission arises in the warmer surface layers, as revealed by X-shaped patterns in their zeroth-moment maps \citep{podio2020,vanthoff2020}. These observations, however, targeted the \hhco{} $3_{12}-2_{11}$ line which has a relatively large upper-level energy of $\sim33$~K, and is therefore more sensitive to the warmer elevated disk layers. Lower-energy lines could trace colder gas present at intermediate disk layers. In order to confirm the origin of the \hhco{} emission, multi-line observations are needed \citep[e.g.,][]{pegues2020}. The results toward Class II disks found by \citet{pegues2020} suggest a possible trend between the location of the \hhco{} emitting layer and the age of the disk, or perhaps a difference between disks around T~Tauri and Herbig stars. Future multi-line observations toward a larger sample of disks with a consistent set of lines may help to elucidate whether this trend is real.

\subsection{Small organics reservoir in the inner $50$~au}

We can estimate the amount of HCN, \cch{}, and \hhco{} present in the inner disk regions. The gas masses within $50$~au are listed in Table~\ref{tab:organic_reservoir}. Because of the uncertainties of the fit in the inner disk, we extrapolated the column density profiles inward, assuming a constant column density starting from the radius where we obtain a good fit (see~Fig.~\ref{fig:Nprofs}). This means that the HCN masses estimated for GM~Aur, HD~163296, and MWC~480, which tend to have centrally peaked N profiles, could be larger than what we have estimated here. The total amount of HCN, \cch{}, and \hhco{} is, however, much larger because a large fraction of the organic reservoir is expected to reside in icy mantles. Indeed, as discussed before, the emission of these small organics arises in disk layers with temperatures of $20-60$~K, which are lower than the freeze-out temperature of these molecules. Assuming a conservative ice-to-gas ratio of 1000 \citep[these ratios could be larger, as discussed in][]{oberg21}, we find a large amount ($10^{23} - 10^{26}$~g) of gas phase plus ice mantle HCN, \cch{}, and \hhco{} in the inner disks. Interestingly, the relative amounts of these organics varies between the disks. For example, \hhco{} is $\sim5$ times more abundant than HCN and \cch{} in IM~Lup, while \hhco{} is less abundant than HCN and \cch{} in the rest of the disks. HCN is more abundant (by a factor $\sim10-50$) than both \cch{} and \hhco{} in GM~Aur, HD~163296, and MWC~480, and AS~209 has a slightly larger \cch{} mass than HCN.

We can also compare the amount of organics with respect to water ice in these disks. The total water mass inside $50$~au was estimated from the models presented in \citet{zhang21}, and are listed in Table~1 of \citet{oberg21}. The abundance of total gas and ice HCN, \cch{}, and \hhco{} relative to water are listed in the right columns of Table~\ref{tab:organic_reservoir}. The fractions range from 0.5\% to 1.1\% and 0.02\% to 1.8\% for HCN and \cch{}, respectively, not taking into account IM~Lup, which has much lower organic fractions of $<0.001\%$. The fractions for \hhco{} are lower, ranging from 0.001\% to 0.08\%. Typical abundances with respect to water in cometary ices range from 0.1\% to 0.6\% for HCN and from $0.1-1\%$ for \hhco{} \citep{mumma2011}, which are consistent with our organic fraction estimates. 

The MAPS disks are therefore rich in organic species, suggesting that future comets formed in these disks could efficiently deliver water and other key organics to rocky planets forming in the inner disk.

\section{Conclusions} \label{sec:conclusons}

We have presented spatially resolved observations of the small organics HCN, \cch{}, and \hhco{} toward a sample of five protoplanetary disks from the MAPS program. The main conclusions are the following:

\begin{enumerate}
\item The line emission reveals substantial substructure, in particular toward the warmer disks around HD~163296 and MWC~480, where several rings are observed in HCN and \cch{}. 
\item We derive column density profiles and find the distributions vary across disks, suggesting that the specific physical conditions in each disk play a major role in setting the organic distribution in disks. 
\item We find similarities between the HCN and \cch{} distribution in the outer disk regions, suggesting a similar chemistry drives their formation, most likely photochemistry in the temperate disk layers. 
\item The HCN and \cch{} excitation temperatures, estimated by the $1-0$ and $3-2$ lines as well as the hyperfine structure of the lines, range between 20 and 60~K. The temperatures are higher for \cch{}, however, suggesting that the HCN emission arises in a colder and deeper layer than \cch{}, and is thus more sensitive to density variations. 
\item HD~163296 and MWC~480 show gaps in their HCN column density profiles, which spatially coincide with the location of planets suggested in previous studies. This opens the possibility of using HCN to trace gas density variations in disks, and finding the massive planets that could be carving these gaps.
\item Contrary to HCN and \cch{}, \hhco{} emits from a much more extended region of the disks, far beyond the dust millimeter edge. \hhco{} is slightly more abundant in the colder disks around T~Tauri stars than in disks around Herbig stars, but is particularly abundant in the GM~Aur disk. 
\item We find enhanced \hhco{} column densities at the CO snowline location and another column density enhancement around the millimeter dust edge. 
\item We estimated the total gas plus ice mass inside $50$~au for the MAPS disks, and find large amounts ($10^{23}-10^{26}$~g) of organic material, with fractions with respect to water ice that are consistent with fractions observed in comets of our Solar System.
\end{enumerate}

We have shown that high angular resolution (down to $\sim10$~au) and sensitivity observations from these small organics provide a window into the C, O, and N budget of disks. The inner 50~au regions of the MAPS disks are rich in organic species, so comets forming in these disks could potentially deliver water and other key organics to nascent rocky planets. Future observations toward other disks with different stellar masses, ages, and disk properties will provide a more complete picture of the organic reservoir in disks, and the impact these properties may have on the possible outcomes of planet formation.

\acknowledgments

We thank the anonymous referee for valuable comments that improved this manuscript.

This paper makes use of the following ALMA data: ADS/JAO.ALMA\#2018.1.01055.L. ALMA is a partnership of ESO (representing its member states), NSF (USA) and NINS (Japan), together with NRC (Canada), MOST and ASIAA (Taiwan), and KASI (Republic of Korea), in cooperation with the Republic of Chile. The Joint ALMA Observatory is operated by ESO, AUI/NRAO and NAOJ. The National Radio Astronomy Observatory is a facility of the National Science Foundation operated under cooperative agreement by Associated Universities, Inc. 

V.V.G. acknowledges support from FONDECYT Iniciaci\'on 11180904 and ANID project Basal AFB-170002. 
J.B.B. acknowledges support from NASA through the NASA Hubble Fellowship grant \#HST-HF2-51429.001-A, awarded by the Space Telescope Science Institute, which is operated by the Association of Universities for Research in Astronomy, Inc., for NASA, under contract NAS5-26555. 
C.J.L. acknowledges funding from the National Science Foundation Graduate Research Fellowship under Grant DGE1745303. 
K.I.\"O acknowledges support from the Simons Foundation (SCOL \#321183) and an NSF AAG Grant (\#1907653). 
C.W. acknowledges financial support from the University of Leeds, STFC and UKRI (grant numbers ST/R000549/1, ST/T000287/1, MR/T040726/1).
G.C. is supported by NAOJ ALMA Scientific Research Grant Code 2019-13B.
Y.A. acknowledges support by NAOJ ALMA Scientific Research Grant Code 2019-13B, and Grant-in-Aid for Scientific Research 18H05222 and 20H05847. 
E.A.B. acknowledges support from NSF AAG Grant \#1907653. S.M.A. and J.H. acknowledge funding support from the National Aeronautics and Space Administration under Grant No. 17-XRP17 2-0012 issued through the Exoplanets Research Program. 
IC was supported by NASA through the NASA Hubble Fellowship grant HST-HF2-51405.001-A awarded by the Space Telescope Science Institute, which is operated by the Association of Universities for Research in Astronomy, Inc., for NASA, under contract NAS5-26555.  
J.H. acknowledges support for this work provided by NASA through the NASA Hubble Fellowship grant \#HST-HF2-51460.001-A awarded by the Space Telescope Science Institute, which is operated by the Association of Universities for Research in Astronomy, Inc., for NASA, under contract NAS5-26555. 
K.Z. acknowledges the support of the Office of the Vice Chancellor for Research and Graduate Education at the University of Wisconsin–Madison with funding from the Wisconsin Alumni Research Foundation, and support of NASA through Hubble Fellowship grant HST-HF2-51401.001. awarded by the Space Telescope Science Institute, which is operated by the Association of Universities for Research in Astronomy, Inc., for NASA, under contract NAS5-26555. 
R.L.G. acknowledges support from a CNES fellowship grant. 
F.A. acknowledge support from NSF AAG Grant \#1907653.
J.D.I. acknowledges support from the Science and Technology Facilities Council of the United Kingdom (STFC) under ST/T000287/1.
R.T. acknowledges support from the Smithsonian Institution as a Submillimeter Array (SMA) Fellow.
L.I.C. gratefully acknowledges support from the David and Lucille Packard Foundation and Johnson \& Johnson's WiSTEM2D Program.
F.L. acknowledges support from the Smithsonian Institution as a Submillimeter Array (SMA) Fellow.
K.R.S. acknowledges the support of NASA through the NASA Hubble Fellowship grant \#HST-HF2-51419.001 awarded by the Space Telescope Science Institute, which is operated by the Association of Universities for Research in Astronomy, Inc., for NASA, under contract NAS5-26555.
A.D.B acknowledges support from NSF AAG Grant \#1907653.
L.M.P.\ acknowledges support from ANID project Basal AFB-170002 and from ANID FONDECYT Iniciaci\'on project \#11181068. 
F.M. acknowledges support from ANR of France under contract ANR-16-CE31-0013 (Planet-Forming-Disks)  and ANR-15-IDEX-02 (through CDP ``Origins of Life"). 
Y.L. acknowledges the financial support by the Natural Science Foundation of China (Grant No. 11973090).

\software{Astropy \citep{astropy_2013,astropy_2018}, 
    CASA \citep{McMullin_etal_2007}, 
    \texttt{GoFish} \citep{Teague19JOSS}, 
    NumPy \citep{vanderWalt_etal_2011}, 
    SciPy \citep{Virtanen_etal_2020}
    }



\newpage

\appendix
\counterwithin{figure}{section}
\counterwithin{table}{section}

\section{Disk-integrated line fluxes}

The disk-integrated line fluxes and upper limits for the observed HCN, \cch{}, and \hhco{} lines are listed in Table~\ref{tab:fluxes}. For HCN and \cch{}, the integrated line fluxes include several hyperfine components that are blended in velocity. The fluxes were computed using elliptical masks that were constructed using the disk inclinations and position angles listed in Table~\ref{tab:disk_sample}, and an outer radius based on the observed extent of the emission of each line. The uncertainties were estimated by computing integrated fluxes using the same elliptical mask but centered at 500 random positions that are free from line emission in the map. The listed uncertainties correspond to the standard deviation of these line-free integrated fluxes.

\TabFluxes{}

\section{Simultaneous fit to the $3-2$ and $1-0$ lines}
\label{app:fit}

Figure~\ref{fig:NprofsCombined} shows the HCN and \cch{} column density and excitation temperature profiles found when fitting the $1-0$ and $3-2$ lines simultaneously. The spatial resolution of the profiles is 0\farcs3 which corresponds to the resolution of the tapered version of the $1-0$ line images. An example with the resulting best-fit spectra found for AS~209 are shown in Fig.~\ref{fig:spec-as209}. The weakest \cch{} $3-2$ hyperfine line listed in Table~\ref{tab:spec_params} and shown in the bottom panel of the figure is detected in all disks except IM~Lup. 

\FigNprofsCombined{}

\FigSpecLowResAS{}

\section{Comparison of HCN column density profiles}
\label{app:HCNcomparison}

The MAPS data were used to derive HCN column density profiles in two other papers. \citet{bergner21} derived CN and HCN column density profiles using a column density retrieval method similar to the one presented here. \citet{Cataldi21} also derived HCN column density profiles, but using the lower angular resolution images because the goal was to compare them to the profiles of DCN. The Band 6 DCN line is faint and was not detected at high S/N in the high angular resolution images. The resulting HCN column density profiles from these different analyses are shown in Fig.~\ref{fig:Nprofs-compHCN}. The profiles are consistent for all disks, and differences are only observed in the inner disk where the fit is known to behave poorly.

\FigNprofsComparisonHCN{}


\bibliography{references}{}
\bibliographystyle{aasjournal}

\end{document}